\documentclass[12pt]{iopart}
\usepackage{amssymb}

\font\eightrm=cmr8 
\newif\ifinexp \inexpfalse
\newcommand\I{\ifinexp \hbox{\hskip0.8pt\eightrm i} \else
\hbox{\hskip1pt\rm i} \fi}
\newcommand\E[1]{\inexptrue \hbox{e}^{#1} \inexpfalse}

\newcommand\innerp[3]{\pi_{#1,#2}^{(#3)}}
\newcommand\EE[2]{E^{(#1)}_{#2}}
\newcommand\HH[1]{\vphantom{\vbox to #1pt{}}\displaystyle}
\newcommand\hr{\hbox{\bf R}}
\newcommand\hc{\hbox{\bf C}}
\newcommand\hz{\hbox{\bf Z}}
\newcommand\DDDD{{\cal D}}
\newcommand\FFFF{{\cal F}}
\newcommand\MMMM{{\cal M}}
\newcommand\NNNN{{\cal N}}
\newcommand\gggg{{\bf g}}

\newcommand\ad{\,\hbox{\rm ad}}
\newcommand\rank{\hbox{\rm rank}}
\newcommand\res{\mathop{\hbox{\rm Res}}}
\newcommand\D{\displaystyle}
\renewcommand\d{\hbox{\rm d}}
\newcommand\mod{\;\hbox{\rm mod}\;}

\newtheorem{lemma}{\bf Lemma}
\newtheorem{theorem}{\bf Theorem}
\newtheorem{cor}{\bf Corollary}

\newtheorem{remark}{\bf Remark}

\begin{document}
\title[Finite dimensional Hamiltonian system related
to Lax pair with symmetries]{Finite dimensional Hamiltonian system
related to Lax pair with symplectic and cyclic symmetries}

\author{Zi-Xiang Zhou}
\address{School of Mathematical Sciences, Fudan University, Shanghai
200433, China} \ead{zxzhou@fudan.edu.cn}

% keyword:
% Lax pair with symmetries,
% Integrable Hamiltonian systems,
% Nonlinear constraints,
% 2+1 dimensional Toda equation

%PACS:
% 02.30.Ik Integrable systems
% 05.45.Yv Solitons
% 02.30.Jr Partial differential equations

% MSC:
% 37K10 Completely integrable systems, integrability tests, bi-Hamiltonian
% structures, hierarchies (KdV, KP, Toda, etc.)
% 70H06 Completely integrable systems and methods of integration
% 35Q51 Soliton-like equations

\begin{abstract}
For the 1+1 dimensional Lax pair with a symplectic symmetry and
cyclic symmetries, it is shown that there is a natural finite
dimensional Hamiltonian system related to it by presenting a unified
Lax matrix. The Liouville integrability of the derived finite
dimensional Hamiltonian systems is proved in a unified way. Any
solution of these Hamiltonian systems gives a solution of the
original PDE. As an application, the two dimensional hyperbolic
$C_n^{(1)}$ Toda equation is considered and the finite dimensional
integrable Hamiltonian system related to it is obtained from the
general results.
\end{abstract}

\section{Introduction}

There are many integrable nonlinear PDEs in 1+1 dimensions
\cite{bib:Babelonbook,bib:Rogersbook}. For quite a few of them, the
related finite dimensional Liouville integrable Hamiltonian systems
have been obtained. By this nonlinearization
method~\cite{bib:Cao0,bib:Cao}, the nonlinear PDE is changed to a
system of nonlinear ODEs which are Liouville integrable Hamiltonian
systems. Any solution of this system of nonlinear ODEs gives a
solution of the original nonlinear PDE. This greatly simplifies the
original problem. It is an effective way to obtain interesting exact
solutions, especially quasi-periodic solutions of the nonlinear
PDEs~\cite{bib:CaoKP,bib:CaoToda1,bib:Qiao,bib:RagCao,bib:Ragnisco1,bib:Ragnisco}.
Soliton solutions can be obtained in this way by a limiting
process~\cite{bib:Zeng}. Some integrable systems in higher
dimensions have also been reduced to finite dimensional Liouville
integrable Hamiltonian
systems~\cite{bib:CaoKP,bib:ChengLi,bib:Kono,bib:ZhouNNV,bib:ZMZ}.

Usually these finite dimensional Hamiltonian systems have Lax
matrices so that the Liouville integrability can be
guaranteed~\cite{bib:Ma,bib:MaZhou,bib:QZZ,bib:ZRG,bib:ZRGNLS}. Most
results are obtained for specific nonlinear PDEs and specific
hierarchies with less symmetries, and the integrability of the
derived finite dimensional Hamiltonian systems are proved case by case.

In the present paper, we consider a quite general Lax pair with a
symplectic symmetry and cyclic symmetries. The Lax matrix is
presented so that the nonlinear constraint of the lowest order is
generated naturally from this Lax matrix. The Hamiltonian function
for the ODEs derived from the nonlinear constraint is expressed in
terms of the Lax matrix. The Liouville integrability of this
Hamiltonian system is proved by obtaining the $r$ matrix and finding
enough functionally independent conserved integrals. This system
contains some known examples such as the MKdV equation and the
nonlinear Schr\"odinger equation. It also contains any $n\times n$
AKNS system with $u(n)$ symmetry, where the symplectic structure is
naturally derived from the complex structure, and the binary
nonlinearization method~\cite{bib:Ma} is recovered. As an
application, the general results are used for the two dimensional
$C_n^{(1)}$ hyperbolic Toda equation~\cite{bib:McIntosh}, one of the
two dimensional affine Toda equations which are all
integrable~\cite{bib:Adler,bib:CaoToda1,bib:Mackay,bib:McIntosh,
bib:Mikhailov2,bib:Nimmo,bib:Nirov1,bib:Terng}. The two dimensional
$C_n^{(1)}$ hyperbolic Toda equation has a natural symplectic
structure. The finite dimensional Hamiltonian systems related to it
are constructed explicitly. These Hamiltonian systems are simpler
than (with space of lower dimension) that presented
in~\cite{bib:ZhouHam} where binary nonlinear constraint was
constructed. The result for the $x$-part of the Lax pair is derived
from the general result of this paper, while that for the $t$-part
which has the $\lambda^{-1}$ term is obtained independently.

The paper is organized as follows. In Section~\ref{sect:LP}, some
notations and the Lax pair with a symplectic symmetry and cyclic
symmetries are presented. In Section~\ref{sect:NC}, the Lax matrix
and nonlinear constraint are obtained for this general system. The
Hamiltonian function is also presented. The $r$ matrix is obtained
in Section~\ref{sect:rmx}, which gives the involution of conserved
integrals. The independence of the conserved integrals which are
enough for Liouville integrability is proved in
Section~\ref{sect:indep}. In Section~\ref{sect:knownegs}, the
specific results, most of which are known, for the $2\times 2$ real
AKNS system, the MKdV equation, the nonlinear Schr\"odinger
equation, the $u(n)$ AKNS system and the $n$ wave equation are
derived from the general conclusions. Finally, in
Section~\ref{sect:Cn}, the results for the two dimensional
$C_n^{(1)}$ Toda equation are derived.

\section{Notations and the Lax pair with symmetries}\label{sect:LP}

Let $W$ be a $2n\times 2n$ invertible antisymmetric real matrix
which gives a symplectic structure on $\hr^{2n}$.

Let
\begin{equation}
   G=\{A\in GL(2n,\hc)\,|\,A^TWA=W\},\label{eq:group}
\end{equation}
which is isomorphic to $Sp(n,\hc)$, the complex symplectic algebra.
The inner auto\-morphism group of $G$ is $G/\{\pm I\}$. Let
$p:\,G\to G/\{\pm I\}$ be the natural projection. Let
\begin{equation}
   \gggg=\{X\in gl(2n,\hc)\,|\,X^T=-WXW^{-1}\}
\end{equation}
be the Lie algebra of $G$.

Let $G_0$ be a finite subgroup of $G$ such that each of its element
$A$ satisfies $\bar AA=\pm I$. Here $\bar A$ is the complex
conjugation (without transpose) of $A$.

\begin{lemma}\label{lemma:OmegaOmega}
$p(G_0)$ is a finite Abelian subgroup of $G/\{\pm I\}$. Therefore,
for any $A,B\in G_0$, either $BA=AB$ or $BA=-AB$ holds.
\end{lemma}

\begin{demo}
For any $A,B\in G_0$, $AB^{-1}\in G_0$. Hence $A^{-1}
BAB^{-1}=\pm\overline{AB^{-1}}AB^{-1}=\pm I$, which implies $BA=\pm
AB$. The lemma is proved.
\end{demo}

Suppose $\Omega_1,\cdots,\Omega_N\in G_0$ so that $p(\Omega_a)$
$(a=1,\cdots,N)$ are generators of $p(G_0)$ and suppose the order of
$p(\Omega_a)\in p(G_0)$ is $m_a$. Then, $\Omega_a$'s satisfy
\begin{equation}
   \Omega_a^TW\Omega_a=W,\quad
   \bar\Omega_a=\pm\Omega_a^{-1},\quad \Omega_a^{m_a}=\pm I.
   \label{eq:rel_Omega}
\end{equation}
Let
$\Sigma=\{(\alpha_1,\cdots,\alpha_N)\,|\,\alpha_a\in\hz\,(a=1,\cdots,N)\}$,
$\Sigma_0=\{\alpha=(\alpha_1,\cdots,\alpha_N)\in\Sigma\,|\,0\le\alpha_a<
m_a\,(a=1,\cdots,N)\}$, then we can write
$\Omega^\alpha=\Omega_1^{\alpha_1}\cdots\Omega_N^{\alpha_N}$ etc.
for multi-index $\alpha=(\alpha_1,\cdots,\alpha_N)\in\Sigma$. Denote
$m_0$ to be the exponent of $p(G_0)$, which is the minimal common
multiple of $m_1,\cdots,m_N$.

Let $\omega:\,G_0\to S^1=\{z\in\hc\,|\,|z|=1\}$ be a group
homomorphism such that $\omega(\pm I)=1$. For any $a=1,\cdots,N$,
denote $\omega_a=\omega(\Omega_a)$, then $\omega_a^{m_a}=1$.

For any fixed integer $k$, denote
\begin{equation}
  \D\DDDD_k=\{X\in\gggg\,|\,\bar X=X,\,\Omega X\Omega^{-1}=\omega(\Omega)^kX
  \hbox{ for any }\Omega\in G_0\},
\end{equation}
then $[\DDDD_j,\DDDD_k]\subset\DDDD_{j+k}$. Moreover, if
$X\in\DDDD_k$, then $X^{2j-1}\in\DDDD_{(2j-1)k}$ for any positive
integer $j$. Let $\D\DDDD=\sum_{k=0}^\infty\DDDD_k$, which is a real
Lie subalgebra of $\gggg$.

Denote $G_0\otimes S^1=\{cg\,|\,c\in S^1,g\in G_0\}$. For given
integer $h$, denote
\begin{equation}
   \begin{array}{l}
   \D\Theta_h=\{\theta\in G_0\otimes S^1|\,\bar\theta=\theta,\widetilde\omega(\theta)=1,
   \theta^T=W\theta W^{-1},\\
   \D\qquad\hbox{ and }
   \Omega\theta\Omega^{-1}=\omega(\Omega)^h\theta\hbox{ for any }\Omega\in G_0\}.
   \end{array}\label{eq:defTheta}
\end{equation}
Here $\widetilde\omega:G_0\otimes S^1\to S^1$ is defined as
$\widetilde\omega(cg)=\omega(g)$ for any $g\in G_0$ and $c\in S^1$.
It is well-defined since $\omega(\pm I)=1$. Moreover,
$\Theta_{h'}=\Theta_h$ if $h'\equiv h\mod m_0$.

$\Theta_h$ may be empty. However, $\Theta_0$ is always non-empty
since $I\in\Theta_0$. For $h\ne 0$, $\Theta_h$ is also useful for
some nonlinear PDEs. (See the example of the nonlinear Schr\"odinger
equation in Section~\ref{eq:NLS}.)

\begin{lemma}\label{lemma:Theta}
(i) $\theta^2=\pm I$ for any $\theta\in\Theta_h$.

(ii) $\Theta_h\ne\emptyset$ only if $2h\equiv 0\mod m_0$.

(iii) $AB=BA$ and $AB\in\Theta_h$ hold for any $A\in\Theta_0$,
$B\in\Theta_h$.

(iv) $\theta X=X\theta$ and $\theta X\in\DDDD_{h+k}$ hold for any
$\theta\in\Theta_h$ and $X\in\DDDD_k$.
\end{lemma}

\begin{demo}
Suppose $\theta=cg$ where $g\in G_0$ and $c\in S^1$, then by
(\ref{eq:group}) and (\ref{eq:defTheta}),
$W=g^TWg=(WgW^{-1})Wg=Wg^2$, which implies $g^2=I$ and then
$\theta^2=c^2I$. Moreover, $\overline{\theta^2}=\theta^2$ and $c\in
S^1$ implies $c^2=\pm 1$. Hence (i) is true.

Following (i), (ii) holds since
$\omega(\Omega)^{2h}\theta^2=\Omega\theta^2\Omega^{-1}=\theta^2$ for
any $\Omega\in G_0$.

Suppose $A\in\Theta_0$, $B\in\Theta_h$, then
$BAB^{-1}=\widetilde\omega(B)^0A=A$ implies $AB=BA$. Then it can be checked
that $AB\in\Theta_h$ by the definition (\ref{eq:defTheta}). This
proves (iii).

Suppose $\theta=cg\in\Theta_h$ where $g\in G_0$ and $c\in S^1$.
Since $\omega(g)=1$, we have $\theta X\theta^{-1}=gXg^{-1}=X$, i.e.
$\theta X=X\theta$ for any $X\in\DDDD_k$. Then $(\theta
X)^T=X^T\theta^T=(-WXW^{-1})(W\theta W^{-1})=-W(\theta X)W^{-1}$.
Moreover, $\Omega\theta X\Omega^{-1}=\omega(\Omega)^{h+k}\theta X$
holds for any $\Omega\in G_0$. This proves (iv). The lemma is
proved.
\end{demo}

For fixed integers $p$ and $h$, let
\begin{equation}\fl
   \begin{array}{l}
   \D\FFFF_{p,h}=\Bigg\{f(\tau)=\sum_{j=1}^s \theta f_{s-j}\tau^{j-1}\,\Bigg|
   s\hbox{ is a positive integer, }\theta\in\Theta_h,\,f_{s-j}\in\hr, \\
   \D\quad\hbox{ and } f_{s-j}\ne 0
   \hbox{ holds only when $j$ is even and }h+j\equiv p+1\mod
   m_0\Bigg\}.
   \end{array}
\end{equation}

Here the necessity of $j$ being even when $f_{s-j}\ne 0$ guarantees
that $K^{j-1}\in\DDDD_{j-1}$ when $K\in\DDDD_1$.

Note that $\FFFF_{p',h'}=\FFFF_{p,h}$ if $p'\equiv p\mod m_0$ and
$h'\equiv h\mod m_0$.
\begin{lemma}
(i) $p-h$ must be odd if $m_0$ is even and $\FFFF_{p,h}\ne\{0\}$.

(ii) $[\FFFF_{1,0},\FFFF_{p,h}]=0$ holds for any integers $p$ and
$h$.

(iii) If $f\in\FFFF_{p,h}$, then $f(K)\in\DDDD_p$ when
$K\in\DDDD_1$.
\end{lemma}

\begin{demo}
Suppose $\D f(\tau)=\sum_{j=1}^s \theta
f_{s-j}\tau^{j-1}\in\FFFF_{p,h}$ and $f\ne 0$. (i) holds since
$p-h\equiv j-1\mod m_0$ and $j-1$ is odd when $f_{s-j}\ne 0$. (ii)
follows from (iii) of Lemma~\ref{lemma:Theta}. Now suppose
$f_{s-j}\ne 0$, then $j$ is even and $K^{j-1}\in\DDDD_{j-1}$ since
$K\in\DDDD_1$. (iv) of Lemma~\ref{lemma:Theta} implies $\theta
f_{s-j}K^{j-1}\in\DDDD_{h+j-1}=\DDDD_p$ by the definition of
$\FFFF_{p,h}$. This proves (iii). The lemma is proved.
\end{demo}

\begin{lemma}\label{lemma:FFFF}
Suppose $f\in\FFFF_{1,0}$, $g\in\FFFF_{p,h}$, then their composition
$g\circ f\in\FFFF_{p,h}$.
\end{lemma}

\begin{demo}
Let
\begin{equation}
   f(\tau)=\sum_{j=1}^s\theta_1f_{s-j}\tau^{j-1},\quad
   g(\tau)=\sum_{k=1}^t\theta_2g_{t-k}\tau^{k-1}
\end{equation}
where $\theta_1\in\Theta_0$, $\theta_2\in\Theta_h$, $f_{s-j}\ne 0$ only if
$j$ is even and $j\equiv 2 \mod m_0$, and $g_{t-k}\ne 0$ only if $k$
is even and $k\equiv p+1-h\mod m_0$. Then
\begin{equation}
   \begin{array}{l}
   \D g(f(\tau))=\sum_{k=1}^t\theta_2g_{t-k}
   \Bigg(\sum_{j=1}^s\theta_1f_{s-j}\tau^{j-1}\Bigg)^{k-1}\\
   \D=\sum_{k=1}^t\sum_{j_1=1}^s\cdots\sum_{j_{k-1}=1}^s\theta_2\theta_1^{k-1}f_{s-j_1}\cdots
   f_{s-j_{k-1}}g_{t-k}\tau^{(j_1-1)+\cdots(j_{k-1}-1)}.
   \end{array}
\end{equation}
A term in the above summation is nonzero only if
$k,j_1,\cdots,j_{k-1}$ are all even, $j_1,\cdots,j_{k-1}\equiv 2\mod
m_0$ and $k\equiv p+1-h\mod m_0$. Then,
$(j_1-1)+\cdots+(j_{k-1}-1)+1$ is even and
$(j_1-1)+\cdots+(j_{k-1}-1)+1\equiv p+1-h\mod m_0$. Moreover, (iii)
of Lemma~\ref{lemma:Theta} implies that $\theta_2\theta_1^{k-1}\in\Theta_h$.
Hence $g(f(\tau))\in\FFFF_{p,h}$.
\end{demo}

The space $\FFFF_{p,h}$ will be used in constructing nonlinear
constraint in the next section.

In this paper, we will consider the linear system
\begin{equation}
   \Phi_x=U(x,\lambda)\Phi\label{eq:LP}
\end{equation}
where
\begin{equation}
   U(x,\lambda)=\sum_{j=0}^p U_j(x)\lambda^{p-j}
\end{equation}
with $U_j\in\DDDD_{p-j}$ ($j=0,1,\cdots,p$). Equivalently,
$U(x,\lambda)$ satisfies
\begin{equation}
   \begin{array}{l}
   \overline{U(\lambda)}=U(\bar\lambda),\quad
   U(\lambda)^T=-WU(\lambda)W^{-1},\\
   \Omega_aU(\lambda)\Omega_a^{-1}=U(\omega_a\lambda)
   \quad(a=1,\cdots,N).
   \end{array}
   \label{eq:Ured}
\end{equation}
Here the first equation in (\ref{eq:Ured}) means that the
coefficients of $\lambda$ in $U(\lambda)$ are real. The second
equation and the third equation mean that $U(\lambda)$ satisfies a
symplectic symmetry and cyclic symmetries respectively.

The linear system (\ref{eq:LP}) with symmetries (\ref{eq:Ured})
consists of many Lax pairs in 1+1 dimensions. We will consider this
general system in the following Section~\ref{sect:NC},
\ref{sect:rmx} and \ref{sect:indep}. The general results to this
linear system can be used for some specific integrable systems which
will be shown in Section~\ref{sect:knownegs} and \ref{sect:Cn}.

\section{Lax matrix and nonlinear
constraint}\label{sect:NC}

Let $\lambda_1,\cdots,\lambda_r$ be non-zero real numbers such that
$\lambda_j^2$'s are distinct. For $\sigma=1,\cdots,r$, let
$\Phi_\sigma=(\phi_{1\sigma},\cdots,\phi_{2n,\sigma})^T$ be a real
column solution of the linear system
\begin{equation}
   \Phi_{\sigma,x}=U(x,\lambda_\sigma)\Phi_\sigma.\label{eq:LP2}
\end{equation}

We will construct a finite dimensional Lax matrix first. For given
$K\in\DDDD_1$, let
\begin{equation}
   \D L(\lambda)=K+\kappa\sum_{\alpha\in \Sigma_0 }\sum_{\sigma=1}^r
   \frac{\Omega^\alpha \Phi_\sigma
   \Phi_\sigma^T(\Omega^{\alpha})^TW}
   {\lambda-\omega^\alpha\lambda_\sigma}
   \label{eq:Laxop}
\end{equation}
where $\kappa$ is a real constant. This construction has already
been used in \cite{bib:ZRG,bib:ZhouHam} and is similar to those used
in constructing Darboux transformations
\cite{bib:Mikhailov,bib:Nirov1,bib:ZhouToda}.

\begin{lemma}\label{lemma:symA}
$L(\lambda)$ satisfies
\begin{eqnarray}
   &&\D L(\bar\lambda)=\overline{L(\lambda)},\label{eq:LsymAreal}\\
   &&\D (L(\lambda))^T=-WL(\lambda)W^{-1},\label{eq:LsymAsymplectic}\\
   &&\D \Omega_a L(\lambda)\Omega_a^{-1}=\omega_aL(\omega_a\lambda),
   \quad a=1,2,\cdots,N.\label{eq:LsymAcyclic}
\end{eqnarray}
\end{lemma}

\begin{demo}
Owing to (\ref{eq:rel_Omega}), suppose
$\bar\Omega_a=\varepsilon_a\Omega_a^{-1}$ with $\varepsilon_a=\pm
1$. (\ref{eq:LsymAreal}) holds since
$\overline{\Omega^\alpha}=\varepsilon^\alpha\Omega^{-\alpha}$,
$\overline{\omega^\alpha}=\omega^{-\alpha}$, and
$\varepsilon^{2\alpha}=1$.

With $W^T=-W$, (\ref{eq:LsymAsymplectic}) follows from
\begin{equation}
   \begin{array}{rl}
   (L(\lambda))^T&\D=K^T-\kappa\sum_{\alpha\in \Sigma_0
   }\sum_{\sigma=1}^r
   \frac{W\Omega^{\alpha}\Phi_\sigma
   \Phi_\sigma^T(\Omega^{\alpha})^T}
   {\lambda-\omega^{\alpha}\lambda_\sigma}
   \\
   &\D=-WKW^{-1}-\kappa W\sum_{\alpha\in \Sigma_0 }\sum_{\sigma=1}^r
   \frac{\Omega^{\alpha}\Phi_\sigma
   \Phi_\sigma^T(\Omega^{\alpha})^T W}
   {\lambda-\omega^{\alpha}\lambda_\sigma}W^{-1}\\
   &=-WL(\lambda)W^{-1}.
   \end{array}
\end{equation}

To prove (\ref{eq:LsymAcyclic}), we have
\begin{equation}
   \begin{array}{rl}
   L(\omega_a\lambda)&\D=K+\kappa\sum_{\alpha\in \Sigma_0
   }\sum_{\sigma=1}^r
   \frac{\Omega^\alpha \Phi_\sigma
   \Phi_\sigma^T(\Omega^{\alpha})^TW}{\omega_a\lambda-\omega^\alpha\lambda_\sigma}\\
   &\D=K+\omega_a^{-1}\kappa\Omega_a\sum_{\alpha\in \Sigma_0
   }\sum_{\sigma=1}^r
   \frac{\Omega^{\alpha}\Phi_\sigma
   \Phi_\sigma^T(\Omega^{\alpha})^T\Omega_a^TW\Omega_a}
   {\lambda-\omega^\alpha\lambda_\sigma}\Omega_a^{-1}\\
   &=\omega_a^{-1}\Omega_a L(\lambda)\Omega_a^{-1}.
   \end{array}\label{eq:LLomega}
\end{equation}
Here we have shifted $\alpha_a$ to $\alpha_a+1$ at the second
equality in (\ref{eq:LLomega}). The last equality follows from
$\Omega_a K\Omega_a^{-1}=\omega_a K$ and $\Omega_a^TW\Omega_a=W$.
The lemma is proved.
\end{demo}

By Lemma~\ref{lemma:symA}, $L(\lambda)\in\DDDD$ for any
$\lambda\in\hr$. Moreover, if $L(\lambda)$ is expanded as $\D
L(\lambda)=\sum_{j=0}^\infty\lambda^{-j}L_j$ with $L_0=K$, then
$L_j\in\DDDD_{1-j}$.

\begin{cor}\label{cor:symA}
Suppose $f\in\FFFF_{p,h}$ $(p\ge 1)$, then $f(L(\lambda))$ satisfies
\begin{eqnarray}
   &&\D f(L(\bar\lambda))=\overline{f(L(\lambda))},\label{eq:fLsymAreal}\\
   &&\D f(L(\lambda))^T=-Wf(L(\lambda))W^{-1},\label{eq:fLsymAsymplectic}\\
   &&\D \Omega_a f(L(\lambda))\Omega_a^{-1}=\omega_a^pf(L(\omega_a\lambda)),
   \quad a=1,2,\cdots,N.\label{eq:fLsymAcyclic}
\end{eqnarray}
\end{cor}

\begin{demo}
Suppose $\D f(\tau)=\sum_{j=1}^s\theta f_{s-j}\tau^{j-1}$ where
$\theta\in\Theta_h$, $f_{s-j}\in\hr$, then we can check that $\D
f(L(\lambda))=\sum_{j=1}^s\theta f_{s-j}\big(L(\lambda)\big)^{j-1}$
satisfies (\ref{eq:fLsymAreal})--(\ref{eq:fLsymAcyclic}) by
Lemma~\ref{lemma:symA} and the definition of $\Theta_h$ and
$\FFFF_{p,h}$. This proves the corollary.
\end{demo}

For a Laurent series $\D N(\lambda)=\sum_{j=-\infty}^nN_j\lambda^j$,
define
\begin{equation}
   N(\lambda)_+=\sum_{j=0}^nN_j\lambda^j,\quad
   N(\lambda)_-=\sum_{j=-\infty}^{-1}N_j\lambda^j.
\end{equation}

Write $M(\lambda)=f(L(\lambda))$ and expand it as
\begin{equation}
   M(\lambda)=\sum_{j=0}^\infty M_j\lambda^{-j} \label{eq:M}
\end{equation}
with $M_0=f(K)$. Corollary~\ref{cor:symA} implies that
$M_j\in\DDDD_{p-j}$ if $f\in\FFFF_{p,h}$.

\begin{theorem}\label{thm:Laxop}
Suppose $f\in\FFFF_{p,h}$ $(p\ge 1)$, then $L(\lambda)$ satisfies
\begin{equation}
   L(\lambda)_x=[U(\lambda),L(\lambda)] \label{eq:Lx}
\end{equation}
under the constraint $U(\lambda)=\widetilde U(\lambda)$ where
\begin{equation}
   \widetilde U(\lambda)=\big(\lambda^pf(L(\lambda))\big)_+.\label{eq:Pexpr}
\end{equation}
Moreover, $\D\widetilde U(\lambda)=\sum_{j=0}^p\widetilde U_j\lambda^{p-j}$
satisfies $\widetilde U_j\in\DDDD_{p-j}$.
\end{theorem}

\begin{demo}
By using (\ref{eq:rel_Omega}), (\ref{eq:Ured}) and (\ref{eq:Laxop}),
\begin{equation}
   \begin{array}{l}
   (\Omega^\alpha \Phi_\sigma \Phi_\sigma^T(\Omega^{\alpha})^TW)_x
   =[U(\omega^\alpha\lambda_\sigma),\Omega^\alpha \Phi_\sigma
   \Phi_\sigma^T(\Omega^{\alpha})^TW].
   \end{array}\label{eq:monox}
\end{equation}
Hence
\begin{equation}
   \begin{array}{l}
   \D -L(\lambda)_x+[U(\lambda),L(\lambda)]=-\sum_{j=0}^p[K,\lambda^{p-j}U_j]\\
   \D+\kappa\sum_{\alpha\in\Sigma_0}\sum_{\sigma=1}^r\sum_{j=0}^p
   \frac{\lambda^{p-j}-(\omega^\alpha\lambda_\sigma)^{p-j}}
   {\lambda-\omega^\alpha\lambda_\sigma}
   \Big[U_j,\Omega^\alpha \Phi_\sigma
   \Phi_\sigma^T(\Omega^{\alpha})^TW\Big]
   \end{array}\label{eq:LxUL}
\end{equation}
is a polynomial of $\lambda$. On the other hand, since
$[f(L(\lambda)),L(\lambda)]=0$, we have
\begin{equation}
   \begin{array}{l}
   \big(-L(\lambda)_x+[\widetilde U(\lambda),L(\lambda)]\big)_+
   =\Big[\big(\lambda^pf(L(\lambda))\big)_+,L(\lambda)\Big]_+\\
   =-\Big[\big(\lambda^pf(L(\lambda))\big)_-,L(\lambda)\Big]_+=0.
   \end{array}
\end{equation}
Hence, $L(\lambda)_x=[U(\lambda),L(\lambda)]$ holds identically.
Moreover, Corollary~\ref{cor:symA} implies that
$\widetilde U_j\in\DDDD_{p-j}$. The theorem is proved.
\end{demo}

Therefore, $L$ satisfies the Lax equation (\ref{eq:Lx}) if
$U(\lambda)$ satisfies the constraint $\D
U(\lambda)=\big(\lambda^pf(L(\lambda))\big)_+$.

With the above constraint, (\ref{eq:LP2}) becomes a system of
nonlinear ODEs
\begin{equation}
   \Phi_{\sigma,x}=\big(\lambda^pf(L(\lambda_\sigma))\big)_+\Phi_\sigma.
   \label{eq:ODE}
\end{equation}

\begin{theorem}\label{thm:Ham}
Suppose $f\in\FFFF_{p,h}$ $(p\ge 1)$, then (\ref{eq:ODE}) is a
Hamiltonian system with the Hamiltonian function
\begin{equation}
   H=\frac 1{2\kappa m_1\cdots m_N}\tr\Big(\res \lambda^pF(L(\lambda))\Big)
   \label{eq:H}
\end{equation}
where $F$ is a matrix-valued polynomial satisfying
$F'(\tau)=f(\tau)$ and $F(0)=0$.
\end{theorem}

\begin{demo}
Expand the Lax matrix $L(\lambda)$ as
\begin{equation}
   L(\lambda)=\sum_{j=0}^\infty\lambda^{-j}L_j
\end{equation}
where
\begin{equation}
   L_0=K,\quad L_j=\kappa\sum_{\alpha\in\Sigma_0}\sum_{\sigma=1}^{r}
   (\omega^\alpha\lambda_\sigma)^{j-1}\Omega^\alpha \Phi_\sigma
   \Phi_\sigma^T(\Omega^{\alpha})^TW\quad (j\ge 1).
\end{equation}
Using the expression $\D f(\tau)=\sum_{l=1}^s\theta
f_{s-l}\tau^{l-1}$, we have $\D F(\tau)=\sum_{l=1}^s\frac 1l\theta
f_{s-l}\tau^l$.

Denote $\hat W=W^{-1}$, then
\begin{equation}
   \begin{array}{l}
   \D 2\kappa m_1\cdots m_N\sum_{k=1}^{2n}\hat W_{jk}
   \frac{\partial H}{\partial\phi_{k\sigma}}\\
   \D=\kappa\sum_{\alpha\in\Sigma_0}\sum_{a,b,k=1}^{2n}\sum_{\mu=1}^{+\infty}
   \res\Big(\lambda^{p-\mu}\big(f(L(\lambda))\big)_{ab}
   (\omega^\alpha\lambda_\sigma)^{\mu-1}\\
   \qquad\cdot \hat W_{jk}\big((\Omega^\alpha \Phi_\sigma)_b((\Omega^\alpha)^TW)_{ka}
   +(\Omega^\alpha)_{bk}(\Phi_\sigma^T(\Omega^\alpha)^TW)_a\big)\Big)\\
   \D=\kappa\sum_{\alpha\in\Sigma_0}\sum_{a,b=1}^{2n}\sum_{\mu=1}^{+\infty}
   \res\Big(\lambda^{p-\mu}\big(f(L(\lambda))\big)_{ab}
   (\omega^\alpha\lambda_\sigma)^{\mu-1}\\
   \qquad\cdot\big((\Omega^\alpha \Phi_\sigma)_b(\Omega^{-\alpha})_{ja}
   +(\Omega^{-\alpha}W^{-1})_{jb}(\Phi_\sigma^T(\Omega^\alpha)^TW)_a\big)\Big)\\
   \D=2\kappa\sum_{\alpha\in\Sigma_0}\sum_{\mu=1}^{+\infty}
   \res\Big(\lambda^{p-\mu}(\omega^\alpha\lambda_\sigma)^{\mu-1}
   \big(\Omega^{-\alpha}f(L(\lambda))\Omega^\alpha \Phi_\sigma\big)_j\Big).
   \end{array}
\end{equation}
Here we have used (\ref{eq:rel_Omega}) and
(\ref{eq:fLsymAsymplectic}). Expand
\begin{equation}
   f(L(\lambda))=\sum_{\nu=0}^\infty M_\nu\lambda^{-\nu}
\end{equation}
as in (\ref{eq:M}), then $M_j\in\DDDD_{p-j}$, and
\begin{equation}
   \begin{array}{l}
   \D 2\kappa m_1\cdots m_N\sum_{k=1}^{2n}\hat W_{jk}
   \frac{\partial H}{\partial\phi_{k\sigma}}\\
   \D=2\kappa\sum_{\alpha\in\Sigma_0}
   \sum_{\mu=1}^{+\infty}\sum_{\nu=0}^{+\infty}
   \res\Big(\lambda^{p-\mu-\nu}(\omega^\alpha\lambda_\sigma)^{\mu-1}
   \big(\Omega^{-\alpha}M_\nu\Omega^\alpha \Phi_\sigma\big)_j\Big)\\
   \D=2\kappa\sum_{\alpha\in\Sigma_0}
   \sum_{\nu=0}^p(\omega^\alpha\lambda_\sigma)^{p-\nu}
   \big(\Omega^{-\alpha}M_\nu\Omega^\alpha \Phi_\sigma\big)_j\\
   \D=2\kappa m_1\cdots m_N\sum_{\nu=0}^p\lambda_\sigma^{p-\nu}
   \big(M_\nu \Phi_\sigma\big)_j\\
   \D=2\kappa m_1\cdots m_N\Big(\Big(\lambda^pf(L(\lambda))\Big)_
   +\Big|_{\lambda=\lambda_\sigma}\Phi_\sigma\Big)_j.
   \end{array}
\end{equation}
Therefore, the Hamiltonian equations given by the Hamiltonian
function (\ref{eq:H}) are just (\ref{eq:ODE}). The theorem is
proved.
\end{demo}

In many concrete integrable systems, say, the nonlinear
Schr\"odinger equation in real form, the condition
$U_{p-1}\in\DDDD_{p-1}\cap(\ker\ad K)^\perp$ is needed where $\perp$
refers to the orthogonal compliment with respect to the Killing form
$\langle\cdot,\cdot\rangle$ of $\gggg$. However, usually
$M_1\in\DDDD_{p-1}\cap(\ker\ad K)^\perp$ is not guaranteed when
$\DDDD_{p-1}\cap\ker\ad K\ne\{0\}$. This problem can be solved with
the help of the following Theorem~\ref{thm:red}. Before that, we
need an algebraic lemma.

\begin{lemma}\label{lemma:det}
Let
\begin{equation}
   \MMMM=\left|\begin{array}{ccccc}
   1&\mu_1&\mu_1^2&\cdots&\mu_1^{2n-1}\\
   \vdots&\vdots&\vdots&&\vdots\\
   1&\mu_n&\mu_n^2&\cdots&\mu_n^{2n-1}\\
   0&1&2\mu_1&\cdots&(2n-1)\mu_1^{2n-2}\\
   \vdots&\vdots&\vdots&&\vdots\\
   0&1&2\mu_n&\cdots&(2n-1)\mu_n^{2n-2}\\
   \end{array}\right|,
\end{equation}
then
\begin{equation}
   \det\MMMM=(-1)^{n(n-1)/2}\prod_{1\le j<k\le n}(\mu_k-\mu_j)^4.
\end{equation}
\end{lemma}

\begin{demo}
Denote $f(x)=(1,x,x^2,\cdots,x^{2n-1})^T$,
\begin{equation}\fl
   \MMMM^{(j,k)}(x)=\det\Bigg(\frac{\d^jf}{\d x^j}(x),f(\mu_2),\cdots,f(\mu_n),
   \frac{\d^kf}{\d x^k}(x),\frac{\d f}{\d x}(\mu_2),\cdots,\frac{\d f}{\d x}(\mu_n)\Bigg).
\end{equation} Then $\MMMM^{(0,1)}(\mu_1)=\MMMM$, $\MMMM^{(0,k)}(\mu_2)=0$,
$\MMMM^{(1,k)}(\mu_2)=0$, and $\MMMM^{(k,k)}(x)=0$ for any $k\ge 1$.
Moreover, we have
\begin{equation}\fl
   \begin{array}{l}
   \D\frac{\d}{\d x}\MMMM^{(0,1)}(x)=\MMMM^{(0,2)}(x),\quad
   \D\frac{\d^2}{\d
   x^2}\MMMM^{(0,1)}(x)=\MMMM^{(0,3)}(x)+\MMMM^{(1,2)}(x),\\
   \D\frac{\d^3}{\d
   x^3}\MMMM^{(0,1)}(x)=\MMMM^{(0,4)}(x)+2\MMMM^{(1,3)}(x).
\end{array}
\end{equation}
This implies $\D\frac{\d^k}{\d x^k}\MMMM^{(0,1)}(\mu_2)=0$ for
$k=0,1,2,3$. Since $\MMMM^{(0,1)}(x)$ is a polynomial of $x$,
$\MMMM$ must be of form $(\mu_2-\mu_1)^4F_1(\mu_1,\cdots,\mu_n)$
where $F_1$ is a polynomial. Owing to the symmetry,
$\D\MMMM=\prod_{1\le j<k\le
n}(\mu_k-\mu_j)^4F_2(\mu_1,\cdots,\mu_n)$ where $F_2$ is another
polynomial. However, regarded as a polynomial of $\mu_1$, $\MMMM$ is
of degree $4n-4$. Hence $F_2$ must be a constant. Comparing the
coefficient of $\D\prod_{k=2}^n\mu_k^{2k-4}$, we get
$F_2=(-1)^{n(n-1)/2}$. The lemma is proved.
\end{demo}

\begin{theorem}\label{thm:red}
Suppose $K\in\DDDD_1$ is diagonalizable, $f\in\FFFF_{p,h}$ $(p\ge
1)$. Expand $M(\lambda)=f(L(\lambda))$ as in (\ref{eq:M}) where
$L(\lambda)$ is given by (\ref{eq:Laxop}). Then there exists a
polynomial $\zeta$ such that $\widetilde
M(\lambda)\equiv\zeta(M(\lambda))= M_0+\lambda^{-1}\widetilde
M_1+o(\lambda^{-1})$ with $\widetilde M_1\in\DDDD_{p-1}\cap(\ker\ad
K)^\perp$ and $M_1-\widetilde M_1\in\DDDD_{p-1}\cap\ker\ad K$.
\end{theorem}

\begin{demo}
Let $K=T\Lambda T^{-1}$ where $\Lambda$ is a complex diagonal matrix
and $T$ is a complex invertible matrix. Let $\widetilde m_0=m_0$ if
$m_0$ is even and $\widetilde m_0=2m_0$ if $m_0$ is odd. Let
$\mu_1,\cdots,\mu_l$ be all the distinct eigenvalues of
$\Lambda^{\widetilde m_0}$. By Lemma~\ref{lemma:det}, there is a
unique complex solution $\zeta_j$ $(j=0,1,\cdots,2l-1)$ of the
linear system
\begin{equation}
   \sum_{k=0}^{2l-1}\zeta_k\mu_j^k=0,\quad
   \widetilde m_0\mu_j\sum_{k=0}^{2l-1}k\zeta_k\mu_j^{k-1}=1\quad(j=1,\cdots,l).
\end{equation}
Then
\begin{equation}
   \sum_{k=0}^{2l-1}\zeta_kK^{k\widetilde m_0}=0,\quad
   \widetilde m_0\sum_{k=0}^{2l-1}k\zeta_kK^{k\widetilde m_0}=I.
\end{equation}
Since $K$ is real and $\zeta_j$'s are unique, $\zeta_j$'s must be real.

Let $\D\zeta(\tau)=\tau-\sum_{k=0}^{2l-1}\zeta_k\tau^{k\widetilde
m_0+1}$, then $\zeta\in\FFFF_{1,0}$ since $\widetilde m_0$ is always
even, and $\zeta(K)=K$.

For any $H\in\ker\ad K$,
\begin{equation}
   \begin{array}{l}
   \D\langle H,\widetilde M-M\rangle=-\sum_{k=0}^{2l-1}\zeta_k\langle
   H,M^{k\widetilde m_0+1}\rangle\\
   \D=-\sum_{k=0}^{2l-1}\zeta_k\Bigg\langle
   H,K^{k\widetilde m_0+1}+\lambda^{-1}\sum_{j=0}^{k\widetilde m_0}K^jM_1K^{k\widetilde m_0-j}
   \Bigg\rangle+o(\lambda^{-1})\\
   \D=-\sum_{k=0}^{2l-1}\langle
   H,\zeta_kK^{k\widetilde m_0+1}+\lambda^{-1}(k\widetilde m_0+1)\zeta_kK^{k\widetilde m_0}M_1
   \rangle+o(\lambda^{-1})\\
   =-\lambda^{-1}\langle H,M_1\rangle+o(\lambda^{-1}).
   \end{array}
\end{equation}
Comparing the coefficients of $\lambda^{-1}$, we have $\langle
H,\widetilde M_1\rangle=0$. Since $\zeta\in\FFFF_{1,0}$,
Lemma~\ref{lemma:FFFF} and Corollary~\ref{cor:symA} imply that
$\widetilde M_1\in\DDDD_{p-1}$. Hence $\widetilde
M_1\in\DDDD_{p-1}\cap(\ker\ad K)^\perp$.

On the other hand,
\begin{equation}\fl
   \begin{array}{l}
   \D[K,\widetilde M-M]=-\sum_{k=0}^{2l-1}\zeta_k\Bigg[K,K^{k\widetilde m_0+1}
   +\lambda^{-1}\sum_{j=0}^{k\widetilde m_0}K^jM_1K^{k\widetilde m_0-j}\Bigg]+o(\lambda^{-1})\\
   \D=-\lambda^{-1}\sum_{k=0}^{2l-1}[\zeta_kK^{k\widetilde m_0+1},M_1]+o(\lambda^{-1})
   =o(\lambda^{-1}).
   \end{array}
\end{equation}
Comparing the coefficients of $\lambda^{-1}$, we get $M_1-\widetilde
M_1\in\DDDD_{p-1}\cap\ker\ad K$. The theorem is proved.
\end{demo}

Owing to Lemma~\ref{lemma:FFFF} and Theorem~\ref{thm:red}, we can
always want $U_1\in\DDDD_{p-1}\cap(\ker\ad K)^\perp$ if necessary,
by replacing $f$ with $f\circ\zeta$. However, when $\zeta$ is
complicated, the Hamiltonian function need not be calculated from
Theorem~\ref{thm:Ham}. Instead, it is simpler to integrate it from
the Hamiltonian equations directly. In this case,
Theorem~\ref{thm:Ham} is still important because it shows that the
Hamiltonian function is expressed by the Lax matrix, which is
essential in the proof of Liouville integrability.

\section{$r$ matrix}\label{sect:rmx}

In $\hr^{2n\times r}$ with coordinates $\phi_{j\sigma}$
$(j=1,\cdots,2n$; $\sigma=1,\cdots,r)$, define the symplectic form
\begin{equation}
   \sum_{j,k=1}^{2n}\sum_{\sigma=1}^r
   W_{jk}\d\phi_{j\sigma}\wedge\d\phi_{k\sigma}.
\end{equation}
Then, for any two smooth functions $f$ and $g$, their Poisson
bracket is
\begin{equation}
   \{f,g\}=\sum_{j,k=1}^{2n}\sum_{\sigma=1}^r\hat W_{jk}
   \frac{\partial f}{\partial\phi_{j\sigma}}\frac{\partial
   g}{\partial\phi_{k\sigma}}
   \label{eq:PB}
\end{equation}
with $\hat W=W^{-1}$.
\begin{theorem}\label{thm:rmx}
For any $\lambda,\mu\in\hc$,
\begin{equation}\fl
   \{L_{ab}(\lambda),L_{cd}(\mu)\}=[r_1(\lambda,\mu),L(\lambda)\otimes I]_{abcd}
   +[r_2(\lambda,\mu),I\otimes L(\mu)]_{abcd}
\end{equation}
holds where the Poisson bracket is given by (\ref{eq:PB}) and
\begin{equation}\fl
   \begin{array}{l}
   \D(r_{1}(\lambda,\mu))_{abcd}
   =\sum_{\gamma\in \Sigma_0 }\frac{\kappa}{\mu-\omega^\gamma\lambda}
   \Big((\Omega^{-\gamma})_{ad}
   (\Omega^{\gamma})_{cb}-(\Omega^{-\gamma} W^{-1})_{ac}
   (W\Omega^{\gamma})_{db}\Big)\\
   \D(r_{2}(\lambda,\mu))_{abcd}=\sum_{\gamma\in \Sigma_0 }
   \frac{\kappa\omega^\gamma}{\mu-\omega^\gamma\lambda}
   \Big((\Omega^{-\gamma})_{ad}
   (\Omega^{\gamma})_{cb}-(\Omega^{-\gamma} W^{-1})_{ac}
   (W\Omega^{\gamma})_{db}\Big)\\
   \qquad=-(r_1(\mu,\lambda))_{cdab}.
   \end{array}
\end{equation}
\end{theorem}
Here
\begin{equation}
   [A,B]_{abcd}=\sum_{p,q=1}^{2n}(A_{apcq}B_{pbqd}-B_{apcq}A_{pbqd})
\end{equation}
for any two $(2n)^2\times(2n)^2$ matrices $A$ and $B$.

\begin{demo}
Written in components,
\begin{equation}\fl
   L_{ab}(\lambda)
   =K_{ab}+\sum_{\alpha\in\Sigma_0}\sum_{\sigma=1}^r\sum_{f,g,h=1}^{2n}
   \frac{\kappa(\Omega^\alpha)_{af}\phi_{f\sigma}
   \phi_{g\sigma}(\Omega^\alpha)_{hg}W_{hb}}
   {\lambda-\omega^\alpha\lambda_\sigma},
\end{equation}
\begin{equation}\fl
   \begin{array}{l}
   \D\frac{\partial L_{ab}(\lambda)}{\partial\phi_{j\sigma}}
   \D=\sum_{\alpha\in\Sigma_0}\sum_{f,h=1}^{2n}\Bigg(
   \frac{\kappa(\Omega^\alpha)_{aj}
   \phi_{f\sigma}(\Omega^\alpha)_{hf}W_{hb}}
   {\lambda-\omega^\alpha\lambda_\sigma}
   +\frac{\kappa(\Omega^\alpha)_{af}\phi_{f\sigma}
   (\Omega^\alpha)_{hj}W_{hb}}
   {\lambda-\omega^\alpha\lambda_\sigma}\Bigg),
   \end{array}
\end{equation}
\begin{equation}\fl
   \begin{array}{l}
   \D\frac{\partial L_{cd}(\mu)}{\partial\phi_{k\sigma}}
   \D=\sum_{\beta\in\Sigma_0}\sum_{q,r=1}^{2n}\Bigg(
   \frac{\kappa(\Omega^\beta)_{ck}
   \phi_{q\sigma}(\Omega^\beta)_{rq}W_{rd}}
   {\mu-\omega^\beta\lambda_\sigma}
   +\frac{\kappa(\Omega^\beta)_{cq}\phi_{q\sigma}
   (\Omega^\beta)_{rk}W_{rd}}
   {\mu-\omega^\beta\lambda_\sigma}\Bigg).
   \end{array}
\end{equation}

The Poisson bracket is
\begin{equation}\fl
   \begin{array}{rl}
   \D\Delta_{abcd}\equiv& \D\{L_{ab}(\lambda),L_{cd}(\mu)\}
   =\sum_{\sigma=1}^r\sum_{j,k=1}^{2n}\hat W_{jk}
   \frac{\partial L_{ab}(\lambda)}{\partial\phi_{j\sigma}}
   \frac{\partial L_{cd}(\mu)}{\partial\phi_{k\sigma}}\\
   \D=&\D\sum_{\sigma=1}^r\sum_{\alpha,\beta\in\Sigma_0}\sum_{j,k,f,h,q,r=1}^{2n}
   \kappa^2
   \D\Big(\frac{\omega^\alpha}{\lambda-\omega^\alpha\lambda_\sigma}
   -\frac{\omega^\beta}{\mu-\omega^\beta\lambda_\sigma}\Big)
   \frac{\phi_{f\sigma}\phi_{q\sigma}}{\omega^\alpha\mu-\omega^\beta\lambda}\hat W_{jk}\\
   &\D\cdot\Big((\Omega^\alpha)_{aj}
   (\Omega^\alpha)_{hf}W_{hb}
   +(\Omega^\alpha)_{af}
   (\Omega^\alpha)_{hj}W_{hb}\Big)\\
   &\D\cdot\Big((\Omega^\beta)_{ck}
   (\Omega^\beta)_{rq}W_{rd}
   +(\Omega^\beta)_{cq}
   (\Omega^\beta)_{rk}W_{rd}\Big)\\
   =&\D{\kappa^2}\sum_{\sigma=1}^r\sum_{\alpha,\beta\in\Sigma_0}
   \Bigg(\frac1{\mu-\omega^{\beta-\alpha}\lambda}
   \frac1{\lambda-\omega^\alpha\lambda_\sigma}
   +\frac1{\lambda-\omega^{\alpha-\beta}\mu}
   \frac1{\mu-\omega^\beta\lambda_\sigma}\Bigg)D_{abcd\alpha\beta}
   \end{array}
\end{equation}
where
\begin{equation}\fl
   \begin{array}{l}
   D_{abcd\alpha\beta}=(\Omega^{\alpha-\beta} W^{-1})_{ac}
   (W^T\Omega^\alpha \Phi_\sigma\Phi_\sigma^T(\Omega^\beta)^TW)_{bd}
   +(\Omega^{\alpha-\beta})_{ad}
   (W^T\Omega^\alpha \Phi_\sigma\Phi_\sigma^T(\Omega^\beta)^T)_{bc}\\
   \D\qquad-(\Omega^{\beta-\alpha})_{cb}
   (\Omega^\alpha \Phi_\sigma\Phi_\sigma^T(\Omega^\beta)^TW)_{ad}
   -(W\Omega^{\alpha-\beta})_{bd}
   (\Omega^\alpha \Phi_\sigma\Phi_\sigma^T(\Omega^\beta)^T)_{ac}.
   \end{array}
\end{equation}
Here we have used (\ref{eq:rel_Omega}).

Let $\gamma=\beta-\alpha$, $\Pi_{\alpha}^{(\sigma)}=\Omega^\alpha
\Phi_\sigma\Phi_\sigma^T(\Omega^\alpha)^TW$, then
$(\Pi_\alpha^{(\sigma)})^T=-W\Pi_\alpha^{(\sigma)}W^{-1}$,
$\Pi_\beta^{(\sigma)}=\Omega^\gamma
\Pi_\alpha^{(\sigma)}\Omega^{-\gamma}$.

Written in terms of $\Pi_\alpha^{(\sigma)}$,
\begin{equation}\fl
   \begin{array}{rl}
   \D D_{abcd\alpha\beta}
   =&\D-(\Omega^{-\gamma} W^{-1})_{ac}(W\Omega^{\gamma}\Pi_\alpha^{(\sigma)})_{db}
   \D+(\Omega^{-\gamma})_{ad}(\Omega^{\gamma}\Pi_\alpha^{(\sigma)})_{cb}\\
   &-(\Omega^{\gamma})_{cb}(\Pi_\alpha^{(\sigma)}\Omega^{-\gamma})_{ad}
   +(W\Omega^{\gamma})_{db}(\Pi_\alpha^{(\sigma)}\Omega^{-\gamma} W^{-1})_{ac}.
   \end{array}
\end{equation}
On the other hand, written in terms of $\Pi_\beta^{(\sigma)}$,
\begin{equation}\fl
   \begin{array}{rl}
   D_{abcd\alpha\beta}
   =&-(\Omega^{-\gamma} W^{-1})_{ac}(W\Omega^{-\gamma}\Pi_\beta^{(\sigma)})_{bd}
   +(\Omega^{-\gamma})_{ad}(\Pi_\beta^{(\sigma)}\Omega^{\gamma})_{cb}\\
   &-(\Omega^{\gamma})_{cb}(\Omega^{-\gamma}\Pi_\beta^{(\sigma)})_{ad}
   +(W\Omega^{\gamma})_{db}(\Pi_\beta^{(\sigma)}\Omega^{\gamma} W^{-1})_{ca}.
   \end{array}
\end{equation}
Hence
\begin{equation}\fl
   \begin{array}{rl}
   \Delta_{abcd}=&\D\sum_{\gamma\in\Sigma_0}\sum_{l=1}^{2n}
   \frac{\kappa}{\mu-\omega^{\gamma}\lambda}\\
   &\D\cdot
   \Big(-(\Omega^{-\gamma} W^{-1})_{ac}(W\Omega^{\gamma}(L(\lambda)-K))_{db}
   +(\Omega^{-\gamma})_{ad}(\Omega^{\gamma}(L(\lambda)-K))_{cb}\\
   &-(\Omega^{\gamma})_{cb}((L(\lambda)-K)\Omega^{-\gamma})_{ad}
   +(W\Omega^{\gamma})_{db}((L(\lambda)-K)\Omega^{-\gamma} W^{-1})_{ac}\Big)\\
   &+\D\sum_{\gamma\in\Sigma_0}\sum_{l=1}^{2n}
   \frac{\kappa\omega^\gamma}{\mu-\omega^{\gamma}\lambda}\\
   &\D\cdot
   \Big(-(\Omega^{-\gamma} W^{-1})_{ac}(W\Omega^{-\gamma}(L(\mu)-K))_{bd}
   +(\Omega^{-\gamma})_{ad}((L(\mu)-K)\Omega^{\gamma})_{cb}\\
   &-(\Omega^{\gamma})_{cb}(\Omega^{-\gamma}(L(\mu)-K))_{ad}
   +(W\Omega^{\gamma})_{db}((L(\mu)-K)\Omega^{\gamma} W^{-1})_{ca}\Big).
   \end{array}\label{eq:Delta}
\end{equation}
In $\Delta_{abcd}$, the terms with $K_{jk}$'s are
\begin{equation}
   \begin{array}{rl}
   &\D\sum_{\gamma\in\Sigma_0}\frac{\kappa}{\mu-\omega^{\gamma}\lambda}
   \Big((\Omega^{-\gamma} W^{-1})_{ac}(W\Omega^{\gamma}K)_{db}
   -(\Omega^{-\gamma})_{ad}(\Omega^{\gamma}K)_{cb}\\
   &+(\Omega^{\gamma})_{cb}(K\Omega^{-\gamma})_{ad}
   -(W\Omega^{\gamma})_{db}(K\Omega^{-\gamma} W^{-1})_{ac}\Big)\\
   &\D+\sum_{\gamma\in\Sigma_0}\frac{\kappa\omega^\gamma}{\mu-\omega^{\gamma}\lambda}
   \Big((\Omega^{-\gamma} W^{-1})_{ac}(W\Omega^{-\gamma}K)_{bd}
   -(\Omega^{-\gamma})_{ad}(K\Omega^{\gamma})_{cb}\\
   &+(\Omega^{\gamma})_{cb}(\Omega^{-\gamma}K)_{ad}
   -(W\Omega^{\gamma})_{db}(K\Omega^{\gamma} W^{-1})_{ac}\Big)=0,
   \end{array}
\end{equation}
in which we have used the relations in (\ref{eq:rel_Omega}) and the
fact $K\in\DDDD_1$. Hence
\begin{equation}\fl
   \begin{array}{rl}
   \D\Delta_{abcd}
   =&\D\sum_{\gamma\in\Sigma_0}\sum_{l=1}^{2n}\frac{\kappa}{\mu-\omega^{\gamma}\lambda}
   \Big(-(\Omega^{-\gamma} W^{-1})_{ac}(W\Omega^{\gamma})_{dl}L_{lb}(\lambda)
   +(\Omega^{-\gamma})_{ad}(\Omega^{\gamma})_{cl}L_{lb}(\lambda)\\
   &-L_{al}(\lambda)(\Omega^{-\gamma})_{ld}(\Omega^{\gamma})_{cb}
   +L_{al}(\lambda)(\Omega^{-\gamma} W^{-1})_{lc}
   (W\Omega^{\gamma})_{db}\Big)\\
   +&\D\sum_{\gamma\in\Sigma_0}\sum_{l=1}^{2n}
   \D\frac{\kappa\omega^{\gamma}}{\mu-\omega^{\gamma}\lambda}
   \Big(-(\Omega^{-\gamma} W^{-1})_{ac}(W\Omega^{\gamma})_{lb}L_{ld}(\mu)
   +(\Omega^{-\gamma})_{al}
   (\Omega^{\gamma})_{cb}L_{ld}(\mu)\\
   &-L_{cl}(\mu)(\Omega^{-\gamma})_{ad}(\Omega^{\gamma})_{lb}
   +L_{cl}(\mu)(\Omega^{-\gamma} W^{-1})_{al}
   (W\Omega^{\gamma})_{db}\Big),
   \end{array}
\end{equation}
which is the result of the theorem.
\end{demo}

From Theorem~\ref{thm:rmx}, it is easy to derive
\begin{theorem}\label{thm:invol}
Suppose $\theta_1$ and $\theta_2$ are constant $2n\times 2n$
matrices such that $[\theta_j,L(\lambda)]=0$ $(j=1,2)$, then
$\{\tr(\theta_1 L(\lambda)^k),\tr(\theta_2 L(\mu)^l)\}=0$ holds for
any positive integers $k$ and $l$, and complex numbers $\lambda$ and
$\mu$.
\end{theorem}

\begin{demo}
According to Theorem~\ref{thm:rmx},
\begin{equation}\fl
   \begin{array}{l}
   \D\quad\frac 1{kl}\{\tr(\theta_1L(\lambda)^k),\tr(\theta_2\theta L(\mu)^l)\}\\
   \D=\sum_{a,b,c,d=1}^{2n}\big(\theta_1L(\lambda)^{k-1}\big)_{ba}
   \big(\theta_2L(\mu)^{l-1}\big)_{dc}
   \big\{L(\lambda)_{ab},L(\mu)_{cd}\big\}\\
   \D=\sum_{a,b,c,d,j=1}^{2n}\big(\theta_1L(\lambda)^{k-1}\big)_{ba}
   \big(\theta_2L(\mu)^{l-1}\big)_{dc}\\
   \D\quad\cdot\Big((r_1)_{ajcd}L(\lambda)_{jb}-L(\lambda)_{aj}(r_1)_{jbcd}
   +(r_2)_{abcj}L(\mu)_{jd}-L(\mu)_{cj}(r_2)_{abjd}\Big)\\
   =0
   \end{array}
\end{equation}
by the relations
\begin{equation}
   \sum_{b=1}^{2n}\big(\theta_1L(\lambda)^{k-1}\big)_{ba}L(\lambda)_{jb}
   =\big(\theta_1L(\lambda)^k\big)_{ja}
\end{equation}
etc. The theorem is proved.
\end{demo}

According to Theorem~\ref{thm:Ham} and \ref{thm:invol},
$\{H,\tr(\theta L(\lambda)^k)\}=0$ holds for any positive integer
$k$, complex number $\lambda$ and matrix $\theta$ with
$[\theta,L(\lambda)]=0$.

\section{Independence of conserved integrals} \label{sect:indep}

According to (\ref{eq:LsymAsymplectic}), $\tr (\theta
L(\lambda)^{2k-1})=0$ for any $K\in\DDDD_1$, $\theta\in\Theta_h$,
and positive integers $k$ and $h$. It is only necessary to consider
$\tr(\theta L(\lambda)^k)$ for even $k$ to generate the conserved
integrals.

For given $\theta\in\Theta_h$, expand
\begin{equation}
   \tr(\theta L(\lambda)^{2k})=\sum_{j=0}^{\infty}s_j^{(2k)}(\theta)\lambda^{-j}.
\end{equation}
By (\ref{eq:LsymAcyclic}),
$s_j^{(2k)}(\theta)=\omega_a^{2k+h-j}s_j^{(2k)}(\theta)$ for all
$a=1,\cdots,N$. Hence $s_j^{(2k)}(\theta)=0$ unless $j\equiv
2k+h\mod m_0$.

To consider the non-zero $s_j^{(2k)}(\theta)$'s, let
\begin{equation}
   \EE kp(\theta)=\frac 1{2k}s_{m_0(p-1)+2k+h}^{(2k)}(\theta).
\end{equation}

\begin{theorem}\label{thm:indep}
Suppose $K\in\DDDD_1$ is diagonalizable. Suppose also that there
exist $\theta_k\in\Theta_{h_k}$ $(k=1,\cdots,n)$ such that
$[\theta_j,\theta_k]=0$ for all $j,k$, and $\theta_jK^{2j-1}$
$(j=1,\cdots,n)$ are linearly independent. Then $\EE kp(\theta_k)$
$(k=1,\cdots,n;p=1,\cdots,r)$ are functionally independent in a
dense open subset of $\hr^{2n\times r}$.
\end{theorem}

\begin{demo}
According to (iv) of Lemma~\ref{lemma:Theta}, $[\theta_j,K]=0$ for
all $j$ since $\theta_j\in\Theta_{h_j}$ and $K\in\DDDD_1$. Moreover,
$\theta_j^2=\pm I$ implies that $\theta_j$'s are diagonalizable.
Hence there exists a $2n\times 2n$ complex invertible matrix $T$
such that $K^{(0)}=TKT^{-1}$ and $\theta_j^{(0)}=T\theta_jT^{-1}$
are all complex diagonal matrices. Let $\Xi=(\xi_{jk})_{1\le j\le
n,1\le k\le 2n}$ where $\xi_{jk}$ is the $(k,k)$ entry of
$\theta_j^{(0)}(K^{(0)})^{2j-1}$. Since $\theta_jK^{2j-1}$
$(j=1,\cdots,n)$ are linearly independent, $\rank(\Xi)=n$. Without
loss of generality, suppose that the first $n$ columns of $\Xi$ are
linearly independent.

Let
\begin{equation}
   S=\{\Psi\in\hr^{2n\times r}\,|\,\hbox{all the entries of $T\Psi$ are
   non-zero}\},
\end{equation}
then $S$ is a dense subset of $\hr^{2n\times r}$.

Now we compute the Jacobian matrix of $\EE kp(\theta_k)$ with respect to
$\phi_{j\sigma}$. By the definition of $L(\lambda)$,
\begin{equation}
   \begin{array}{l}
   \D\frac 1{2k}\tr(\theta_kL(\lambda)^{2k})
   =\frac 1{2k}\tr\Bigg(\theta_k\Big(K+\kappa\sum_{\alpha\in \Sigma_0 }\sum_{\sigma=1}^r
   \frac{\Omega^\alpha \Phi_\sigma
   \Phi_\sigma^T(\Omega^{\alpha})^TW}
   {\lambda-\omega^\alpha\lambda_\sigma}\Big)^{2k}\Bigg)\\
   \D=\frac 1{2k}\tr\Big(\theta_kK^{2k}+2k\kappa\theta_kK^{2k-1}
   \sum_{\alpha\in \Sigma_0 }\sum_{\sigma=1}^r
   \frac{\Omega^\alpha \Phi_\sigma
   \Phi_\sigma^T(\Omega^{\alpha})^TW}
   {\lambda-\omega^\alpha\lambda_\sigma}\Big)+\cdots\\
   \end{array}
\end{equation}
where ``$\cdots$'' represents the terms of $\phi_{j\sigma}$'s whose
degrees are higher than $2$. Hence
\begin{equation}\fl
   \begin{array}{l}
   \D\EE{k}{p}(\theta_k)=
   \kappa\tr\Big(\sum_{\alpha\in \Sigma_0 }\sum_{\sigma=1}^r
   (\omega^\alpha\lambda_\sigma)^{m_0(p-1)+2k+h_k-1}\theta_kK^{2k-1}
   \Omega^\alpha \Phi_\sigma \Phi_\sigma^T(\Omega^{\alpha})^TW\Big)+\cdots\\
   \D=\kappa m_1\cdots m_N\sum_{\sigma=1}^r
   \lambda_\sigma^{m_0(p-1)+2k+h_k-1}\Phi_\sigma^TW\theta_kK^{2k-1}\Phi_\sigma+\cdots
   \end{array}
\end{equation}
since $\omega^{m_0\alpha}=1$, $K\in\DDDD_1$,
$\theta_k\in\Theta_{h_k}$ and (\ref{eq:rel_Omega}) holds.

Denote $\Psi=(\phi_{j\sigma})_{1\le j\le 2n;1\le\sigma\le r}$. For
$k=1,\cdots,n$, $j=1,\cdots,2n$, $p=1,\cdots,r$,
$\sigma=1,\cdots,r$,
\begin{equation}
   \frac{\partial\EE{k}{p}(\theta_k)}{\partial\phi_{j\sigma}}
   =2\kappa m_1\cdots m_N\lambda_\sigma^{m_0(p-1)+2k+h_k-1}
   (W\theta_kK^{2k-1}\Psi)_{j\sigma}
   +\cdots.
\end{equation}
Here we have used the fact that $W\theta_kK^{2k-1}$ is symmetric. Then
\begin{equation}\fl
   \sum_{l=1}^{2n}(W^{-1})_{jl}\frac{\partial\EE{k}{p}(\theta_k)}{\partial\phi_{l\sigma}}
   =2\kappa m_1\cdots m_N
   \lambda_\sigma^{m_0(p-1)+2k+h_k-1}
   (\theta_kK^{2k-1}\Psi)_{j\sigma}
   +\cdots. \label{eq:W-1J}
\end{equation}

Let
\begin{equation}
   \MMMM^{(s)}=(\lambda_\sigma^{m_0(p-1)+2k+h_k-1}(\theta_kK^{2k-1}\Psi)_{j\sigma})_{ns\times
   2ns}\quad(1\le s\le r)
\end{equation}
where the row indices are $k=1,\cdots,n$ and $p=1,\cdots,s$, and the
column indices are $j=1,\cdots,2n$ and $\sigma=1,\cdots,s$. Write
$\MMMM^{(s)}$ as the block matrix
$\MMMM^{(s)}=(\MMMM^{(s)}_{kj})_{1\le k\le n;1\le j\le 2n}$ where
\begin{equation}\fl
   \MMMM^{(s)}_{kj}=\left(\begin{array}{ccc}
   \lambda_1^{2k+h_k-1}(\Psi^{(k)})_{j1}&\cdots
   &\lambda_s^{2k+h_k-1}(\Psi^{(k)})_{js}\\
   \lambda_1^{m_0+2k+h_k-1}(\Psi^{(k)})_{j1}&\cdots
   &\lambda_s^{m_0+2k+h_k-1}(\Psi^{(k)})_{js}\\
   \vdots&&\vdots\\
   \lambda_1^{m_0(s-1)+2k+h_k-1}(\Psi^{(k)})_{j1}&\cdots
   &\lambda_s^{m_0(s-1)+2k+h_k-1}(\Psi^{(k)})_{js}
   \end{array}\right)
\end{equation}
are $s\times s$ matrices, and $\Psi^{(k)}=\theta_kK^{2k-1}\Psi$. Let
\begin{equation}\fl
   \NNNN^{(s)}=\left(\begin{array}{ccc}
   \sigma^{(s)}\\&\ddots\\&&\sigma^{(s)}\end{array}\right)_{ns\times
   ns},\quad
   \sigma^{(s)}=\left(\begin{array}{ccccc}1\\-\lambda^{m_0}_s&1\\
   &-\lambda^{m_0}_s&1\\&&\ddots&\ddots\\&&&-\lambda^{m_0}_s&1
   \end{array}\right)_{s\times s},
\end{equation}
then
\begin{equation}\fl
   \sigma^{(s)}\MMMM^{(s)}_{kj}=\left(\hskip-8pt\begin{array}{cc}
   \begin{array}{ccc}\lambda_1^{2k+h_k-1}(\Psi^{(k)})_{j1}&\cdots
   &\lambda_{s-1}^{2k+h_k-1}(\Psi^{(k)})_{j,s-1}\end{array}
   &\!\!\!\!\lambda_s^{2k+h_k-1}(\Psi^{(k)})_{js}\\
   \begin{array}{ccc}\Big((\lambda^{m_0}_b-\lambda^{m_0}_s)
   \lambda_b^{(a-1)m_0+2k+h_k-1}(\Psi^{(k)})_{jb}\Big)_{1\le
   a,b\le s-1}\end{array}&0_{(s-1)\times 1}
   \end{array}\right)
\end{equation}
and $\MMMM^{(s)}$ is transformed to $\NNNN^{(s)}\MMMM^{(s)}$ under
elementary transformations. Take another elementary transformation
for $\NNNN^{(s)}\MMMM^{(s)}$ by changing the 1st, $(s+1)$-th,
$(2s+1)$-th, $\cdots$, $((n-1)s+1)$-th rows to the bottom and
changing the $s$-th, $2s$-th, $\cdots$, $2ns$-th column to the
right. Then $\MMMM^{(s)}$ is changed to
\begin{equation}
   \left(\begin{array}{cc}\widetilde M^{(s-1)}&0\\**&B_s\end{array}\right)
   \label{eq:MMMMr}
\end{equation}
where $\widetilde\MMMM^{(s-1)}=(\widetilde M^{(s-1)}_{kj})_{1\le
k\le n,1\le j\le 2n}$,
\begin{equation}
   \begin{array}{l}
   \D\widetilde\MMMM_{kj}^{(s-1)}=\MMMM_{kj}^{(s-1)}\left(\begin{array}{ccc}
   \lambda^{m_0}_1-\lambda^{m_0}_s\\&\ddots\\&&\lambda^{m_0}_{s-1}-\lambda^{m_0}_s
   \end{array}\right),\\
   \D B_s=\Bigg(\lambda_s^{2k+h_k-1}(\Psi^{(k)})_{js}\Big)_{1\le k\le
   n,1\le j\le 2n}.
   \end{array}
\end{equation}
Then
\begin{equation}\fl
   \sum_{l=1}^{2n}T_{jl}(B_s)_{kl}=\lambda_s^{2k+h_k-1}(T\theta_kK^{2k-1}\Psi)_{js}
   =\lambda_s^{2k+h_k-1}(\theta^{(0)}_k(K^{(0)})^{2k-1}T\Psi)_{js}.
\end{equation}
That is
\begin{equation}
   (B_sT^T)_{jk}=\lambda_s^{2j+h_j-1}\xi_{jk}(T\Psi)_{ks}.
\end{equation}
Hence
\begin{equation}
   B_sT^T=(\lambda_s^{2j+h_j-1}\delta_{jk})_{n\times n}
   (\xi_{jk})_{n\times 2n}((T\Psi)_{js}\delta_{jk})_{2n\times 2n}
\end{equation}
is of rank $n$ provided that $\Psi\in S$.

Hence $\rank(B_s)=n$ if $\Psi\in S$. From (\ref{eq:MMMMr}), we have
\begin{equation}
   \rank(\MMMM^{(s)})=\rank(\MMMM^{(s-1)})+n
\end{equation}
if $\Psi\in S$, which implies $\rank(\MMMM^{(r)})=nr$ if $\Psi\in
S$.

From (\ref{eq:W-1J}), for given $\Psi_0\in S$, there exists
$\varepsilon_0>0$ such that
\begin{equation}
   \rank\Big(\sum_{l=1}^{2n}(W^{-1})_{jl}
   \frac{\partial\EE{k}{p}(\theta_k)}{\partial\phi_{l\sigma}}\Big)
   _{1\le k\le n,1\le p\le r\atop 1\le j\le 2n,1\le\sigma\le r}
   \Bigg|_{\Psi=\varepsilon\Psi_0}=nr
\end{equation}
holds for all $\varepsilon$ with $|\varepsilon|<\varepsilon_0$.
Equivalently,
$\D\rank\Big(\frac{\partial\EE{k}{p}(\theta_k)}{\partial\phi_{j\sigma}}
\Big)_{1\le k\le n,1\le p\le r\atop 1\le j\le 2n,1\le\sigma\le r}
\Bigg|_{\Psi=\varepsilon\Psi_0}=nr$ holds for all $\varepsilon$ with
$|\varepsilon|<\varepsilon_0$. Because of the real analyticity, the
above equality holds in a dense subset of $\hr^{2n\times r}$. The
theorem is proved.
\end{demo}

Summarizing the results in Theorem~\ref{thm:Ham}, \ref{thm:invol}
and \ref{thm:indep}, we have the final theorem on the integrability.

\begin{theorem}\label{thm:result}
Suppose $K\in\DDDD_1$ is diagonalizable, $f\in\FFFF_{p,h}$ $(p\ge
1)$. Suppose also that there exist $\theta_k\in\Theta_{h_k}$
$(k=1,\cdots,n)$ such that $[\theta_j,\theta_k]=0$ for all $j,k$,
and $\theta_jK^{2j-1}$ $(j=1,\cdots,n)$ are linearly independent.
Then the system (\ref{eq:ODE}) is an integrable Hamiltonian system
in Liouville sense with Hamiltonian function given by (\ref{eq:H}).
\end{theorem}

With stronger conditions on $K$, we have

\begin{cor}\label{cor:result2}
Suppose $K\in\DDDD_1$ is diagonalizable and $K^2$ has at least $n$
distinct eigenvalues. Suppose also that $f\in\FFFF_{p,h}$ $(p\ge
1)$. Then the system (\ref{eq:ODE}) is an integrable Hamiltonian
system in Liouville sense with Hamiltonian function given by
(\ref{eq:H}).
\end{cor}

\begin{demo}
Take $\theta_1=\cdots=\theta_n=I$ in Theorem~\ref{thm:result}. Since
$K^2$ has at least $n$ distinct non-zero eigenvalues, $K$, $K^3$,
$\cdots$, $K^{2n-1}$ are linearly independent. The result follows
from Theorem~\ref{thm:result}.
\end{demo}

\section{Some examples}\label{sect:knownegs}

In this section, we will recover some known results for certain
important integrable equations from the general results in the
present paper. Hereafter, we always write
\begin{equation}
   \innerp jkl=\sum_{\sigma=1}^r\lambda_\sigma^l\phi_{j\sigma}\phi_{k\sigma}.
\end{equation}

\subsection{MKdV equation}

The MKdV equation
\begin{equation}
   u_t+6u^2u_x+u_{xxx}=0
\end{equation}
has the Lax pair
\begin{equation}
   \begin{array}{l}
   \D\Phi_x=\lambda\left(\begin{array}{cc}1&0\\0&-1\end{array}\right)\Phi
   +\left(\begin{array}{cc}0&u\\-u&0\end{array}\right)\Phi,\\
   \D\Phi_t=-4\lambda^3\left(\begin{array}{cc}1&0\\0&-1\end{array}\right)\Phi
   -4\lambda^2\left(\begin{array}{cc}0&u\\-u&0\end{array}\right)\Phi\\
   \D\qquad-2\lambda\left(\begin{array}{cc}u^2&u_x\\u_x&-u^2\end{array}\right)\Phi
   -\left(\begin{array}{cc}0&u_{xx}+2u^3\\-u_{xx}-2u^3&0\end{array}\right)\Phi.
   \end{array}\label{eq:LPMKdV}
\end{equation}

Now $n=1$, $N=1$, $m_1=2$, $\omega_1=-1$, $\D
W=\left(\begin{array}{cc}0&-1\\1&0\end{array}\right)$, $\Omega_1=W$,
$\D K=\left(\begin{array}{cc}1&0\\0&-1\end{array}\right)$. Then
$\D\DDDD_0=\Bigg\{\left(\begin{array}{cc}0&-c\\c&0\end{array}\right)\,
\Bigg|\,c\in\hr\Bigg\}$,
$\D\DDDD_1=\Bigg\{\left(\begin{array}{cc}a&b\\b&-a\end{array}\right)\,
\Bigg|\,a,b\in\hr\Bigg\}$, $\DDDD_0\cap\ker\ad K=\{0\}$,
$\D\Theta_0=\{\pm I_{2\times 2}\}$, $\D\Theta_1=\emptyset$.

Take $\kappa=-1$. Let $\Phi_\sigma$ $(\sigma=1,\cdots,r)$ be column
solutions of (\ref{eq:LPMKdV}) with $\lambda=\lambda_\sigma$. By
(\ref{eq:Laxop}), the Lax matrix is
\begin{equation}
   \begin{array}{l}
   \D L(\lambda)
   \D=\left(\begin{array}{cc}1&0\\0&-1\end{array}\right)
   +\sum_{k=0}^\infty\lambda^{-2k-1}\left(\begin{array}{cc}0&\innerp11{2k}+\innerp22{2k}\\
   -\innerp11{2k}-\innerp22{2k}&0
   \end{array}\right)\\
   \D+\sum_{k=0}^\infty\lambda^{-2k-2}\left(\begin{array}{cc}
   -2\innerp12{2k+1}&\innerp11{2k+1}-\innerp22{2k+1}\\
   \innerp11{2k+1}-\innerp22{2k+1}&2\innerp12{2k+1}\end{array}\right).
   \end{array}
\end{equation}

Take
\begin{equation}
   f^x(\tau)=\tau\in\FFFF_{1,0},\quad
   f^t(\tau)=2(\tau^3-3\tau)\in\FFFF_{3,0}=\FFFF_{1,0},\label{eq:MKdVfs}
\end{equation}
then $f^x(K)=K$, $f^t(K)=-4K$. According to Theorem~\ref{thm:Laxop},
the nonlinear constraint is
\begin{equation}
   u=\innerp 110+\innerp220. \label{eq:MKdVconstr}
\end{equation}
Then
\begin{equation}
   u_x=2(\innerp111-\innerp221),\quad
   u_{xx}=4(\innerp112+\innerp222)+8u\innerp121.
\end{equation}
Under the constraint (\ref{eq:MKdVconstr}), the Lax pair (\ref{eq:LPMKdV}) becomes
\begin{equation}
   \Phi_{\sigma,x}=(\lambda f^x(L(\lambda)))_+|_{\lambda=\lambda_\sigma}\Phi_\sigma,\quad
   \Phi_{\sigma,t}=(\lambda^3f^t(L(\lambda)))_+|_{\lambda=\lambda_\sigma}\Phi_\sigma.
   \label{eq:MKdVLPconstr}
\end{equation}

\begin{remark}
$\D f^x(\tau)=-\frac 12(\tau^3-3\tau)\in\FFFF_{1,0}=\FFFF_{3,0}$
which is proportional to $f^t(\tau)$ will give the same equation as
$f(\tau)=\tau$.
\end{remark}

According to Theorem~\ref{thm:Ham}, the systems in
(\ref{eq:MKdVLPconstr}) are Hamiltonian systems with Hamiltonian
functions
\begin{equation}\fl
   \begin{array}{l}
   \HH{22}H^x=-\frac 18\tr\res\lambda L(\lambda)^2
   =\frac 1{32}\tr\res\lambda\Big(L(\lambda)^4-6L(\lambda)^2\Big)
   =\innerp121+\frac 14(\innerp110+\innerp220)^2,\\
   \HH{22}H^t=-\frac 1{8}\tr\res\lambda^3\Big(L(\lambda)^4-6L(\lambda)^2\Big)
   =-4\innerp123-2(\innerp110+\innerp220)(\innerp112+\innerp222)\\
   \HH{22}\qquad+(\innerp111-\innerp221)^2
   -2(\innerp110+\innerp220)^2\innerp121-\frac 14(\innerp110+\innerp220)^4.
   \end{array}
\end{equation}

These are involutive Hamiltonian systems which are integrable. The
solutions of the corresponding Hamiltonian equations satisfy the
MKdV equation. Therefore, we have recovered some known
results~\cite{bib:ZRG} from our general results.

\subsection{$2\times 2$ real AKNS system}

The $x$-part of the $2\times 2$ real AKNS system is
\begin{equation}
   \Phi_x=\lambda\left(\begin{array}{cc}1&0\\0&-1\end{array}\right)\Phi
   +\left(\begin{array}{cc}0&u\\v&0\end{array}\right)\Phi.\\
   \label{eq:LPAKNS}
\end{equation}

Now $n=1$, $N=0$, $\D
W=\left(\begin{array}{cc}0&-1\\1&0\end{array}\right)$, $\D
K=\left(\begin{array}{cc}1&0\\0&-1\end{array}\right)$. Then
$\D\DDDD_0=\Bigg\{\left(\begin{array}{cc}a&b\\c&-a\end{array}\right)\,
\Bigg|\,a,b,c\in\hr\Bigg\}$, $\DDDD_0\cap\ker\ad
K=\Bigg\{\left(\begin{array}{cc}a\\&-a\end{array}\right)\,
\Bigg|\,a\in\hr\Bigg\}$, $\D\Theta_0=\{\pm I_{2\times 2}\}$.

Take $\kappa=-1$. Let $\Phi_\sigma$ $(\sigma=1,\cdots,r)$ be column
solutions of (\ref{eq:LPAKNS}) with $\lambda=\lambda_\sigma$. The
Lax matrix is
\begin{equation}
   \begin{array}{l}
   \D L(\lambda)
   \D=\left(\begin{array}{cc}1&0\\0&-1\end{array}\right)
   +\sum_{k=0}^\infty\lambda^{-k-1}\left(\begin{array}{cc}-\innerp12k&\innerp11k\\
   -\innerp22k&\innerp12k\end{array}\right).
   \end{array}
\end{equation}

Noticing that $\DDDD_0\cap\ker\ad K\ne\{0\}$, we need to take $\D
f^x(\tau)=-\frac 12(\tau^3-3\tau)\in\FFFF_{3,0}=\FFFF_{1,0}$ as in
Theorem~\ref{thm:red}. According to Theorem~\ref{thm:Laxop}, the
nonlinear constraint is
\begin{equation}
   u=\innerp 110,\quad v=-\innerp220.
\end{equation}
Under this constraint, the Lax pair (\ref{eq:LPAKNS}) becomes
\begin{equation}
   \Phi_{\sigma,x}=(\lambda f^x(L(\lambda)))_+|_{\lambda=\lambda_\sigma}\Phi_\sigma.
   \label{eq:AKNSLPconstr}
\end{equation}

According to Theorem~\ref{thm:Ham} and Corollary~\ref{cor:result2},
the system (\ref{eq:AKNSLPconstr}) is an integrable Hamiltonian
system with Hamiltonian function
\begin{equation}
   \begin{array}{l}
   \HH{22}H^x=\frac 1{16}\tr\res\lambda\Big(L(\lambda)^4-6L(\lambda)^2\Big)
   =\innerp121+\frac 12\innerp110\innerp220.
   \end{array}
\end{equation}
This is a well known result~\cite{bib:Cao}.

\subsection{Nonlinear Schr\"odinger equation}\label{eq:NLS}

The nonlinear Schr\"odinger equation, written in real form, is
\begin{equation}
   \begin{array}{l}
   u_t=v_{xx}+2(u^2+v^2)v,\\
   -v_t=u_{xx}+2(u^2+v^2)u.
   \end{array}
\end{equation}
Denote $\D I=\left(\begin{array}{cc}1\\&1\end{array}\right)$ and $\D
J=\left(\begin{array}{cc}&-1\\1\end{array}\right)$ which play the
role of $1$ and $\I=\sqrt{-1}$ respectively. The Lax pair in real form is
\begin{equation}\fl
   \begin{array}{l}\D\Phi_x=
   \lambda\left(\begin{array}{cc}I\\&-I\end{array}\right)\Phi
   +\left(\begin{array}{cc}&uI+vJ\\-uI+vJ\end{array}\right)\Phi,\\
   \D\Phi_t=-2\lambda^2\left(\begin{array}{cc}J\\&-J\end{array}\right)\Phi
   -2\lambda\left(\begin{array}{cc}&-vI+uJ\\-vI-uJ\end{array}\right)\Phi\\
   \D\qquad-\left(\begin{array}{cc}(u^2+v^2)J&-v_xI+u_xJ\\v_xI+u_xJ&-(u^2+v^2)J
   \end{array}\right)\Phi.
   \end{array}\label{eq:LPNLS}
\end{equation}

Now $n=2$, $N=2$, $m_1=2$, $m_2=2$, $\omega_1=-1$, $\omega_2=1$,
\begin{equation}\fl
   \Omega_1=W=\left(\begin{array}{cccc}0&0&-1&0\\0&0&0&1\\
   1&0&0&0\\0&-1&0&0\end{array}\right),\quad
   \Omega_2=\left(\begin{array}{cc}\I J\\&\I J\end{array}\right),
   \quad
   K=\left(\begin{array}{cc}I\\&-I\end{array}\right).
\end{equation}
Then
\begin{equation}\fl
   \begin{array}{l}
   \D\DDDD_0=\Bigg\{\left(\begin{array}{cc}aJ&bI+cJ\\-bI+cJ&-aJ\end{array}\right)\Bigg|
   a,b,c\in\hr\Bigg\},\\
   \D\DDDD_1=\Bigg\{\left(\begin{array}{cc}
   aI&bI+cJ\\bI-cJ&-aI\end{array}\right)\Bigg|
   a,b,c\in\hr\Bigg\},\\
   \DDDD_0\cap\ker\ad K=
   \Bigg\{\left(\begin{array}{cc}aJ\\&-aJ\end{array}\right)
   \,\Bigg|\,a\in\hr\Bigg\},\\
   \D\Theta_0=\{\pm I_{4\times 4}\},\quad
   \Theta_1=\Bigg\{\pm\left(\begin{array}{cc}J\\&J\end{array}\right)\Bigg\}.
   \end{array}
\end{equation}

Take $\kappa=-1$. Let $\Phi_\sigma$ $(\sigma=1,\cdots,r)$ be column
solutions of (\ref{eq:LPNLS}) with $\lambda=\lambda_\sigma$. The Lax
matrix is
\begin{equation}\fl
   \begin{array}{l}
   \D
   L(\lambda)=\!\!\left(\begin{array}{cc}I\\&-I\end{array}\right)\!\!
   +\!\!\sum_{k=0}^\infty\lambda^{-2k-1}\!\!\left(\begin{array}{cccc}
   -q_1^{(2k)}J&q_2^{(2k)}I+q_3^{(2k)}J\\
   -q_2^{(2k)}I+q_3^{(2k)}J&q_1^{(2k)}J\end{array}\right)\\
   \D+\sum_{k=0}^\infty\lambda^{-2k-2}\!\left(\begin{array}{cccc}
   -q_1^{(2k+1)}I&q_2^{(2k+1)}I+q_3^{(2k+1)}J\\
   q_2^{(2k+1)}I-q_3^{(2k+1)}J&q_1^{(2k+1)}I\end{array}\right)
   \end{array}
\end{equation}
where
\begin{equation}\fl
   \begin{array}{l}
   q_1^{(2k)}=2(\innerp14{2k}+\innerp23{2k}),\quad
   q_2^{(2k)}=\innerp 11{2k}-\innerp 22{2k}+\innerp 33{2k}-\innerp
   44{2k},\\
   q_3^{(2k)}=2(\innerp12{2k}-\innerp34{2k}),\quad
   q_1^{(2k+1)}=2(\innerp13{2k+1}-\innerp24{2k+1}),\\
   q_2^{(2k+1)}=\innerp 11{2k+1}-\innerp 22{2k+1}-\innerp 33{2k+1}+\innerp
   44{2k+1},\quad
   q_3^{(2k+1)}=2(\innerp12{2k+1}+\innerp34{2k+1}).
   \end{array}
\end{equation}

Denote
$\D\theta=\left(\begin{array}{cc}J\\&J\end{array}\right)\in\Theta_1$,
and let
\begin{equation}\fl
    f^x(\tau)=-\frac 12(\tau^3-3\tau)\in\FFFF_{1,0},\quad
    f^t(\tau)=-\frac 14\theta(3\tau^5-10\tau^3+15\tau)\in\FFFF_{2,1},
    \label{eq:NLSfxt}
\end{equation}
then $f^x(K)=K$, $f^t(K)=-2\theta K$.

\begin{remark}
If we take $f^x(\tau)=\tau$, then $f^x(K)=K$, but $L_1$ has a
non-zero projection in $\DDDD_0\cap\ker\ad K\ne\{0\}$. To solve this
problem, we use Theorem~\ref{thm:red} to get $\D\zeta(\tau)=-\frac
12(\tau^3-3\tau)$, which gives $f^x(x)$ in (\ref{eq:NLSfxt}).
\end{remark}

\begin{remark}
$\D f^x(\tau)=\frac 18(3\tau^5-10\tau^3+15\tau)=-\frac 12\theta
f^t(\tau)$ plays the same role as $f^x(\tau)$ in (\ref{eq:NLSfxt})
does.
\end{remark}

According to Theorem~\ref{thm:Laxop}, the nonlinear constraint is
\begin{equation}
   u=\innerp 110-\innerp 220+\innerp 330-\innerp 440,\quad
   v=2(\innerp120-\innerp340). \label{eq:NLSconstr}
\end{equation}
Then
\begin{equation}
   \begin{array}{l}
   u_x=2(\innerp 111-\innerp 221-\innerp 331+\innerp 440)
   -4v(\innerp140+\innerp230),\\
   v_x=4(\innerp121+\innerp341)+4u(\innerp140+\innerp230).
   \end{array}
\end{equation}
Under the constraint (\ref{eq:NLSconstr}), the Lax pair becomes
\begin{equation}
   \Phi_{\sigma,x}=(\lambda f^x(L(\lambda)))_+|_{\lambda=\lambda_\sigma}\Phi_\sigma,\quad
   \Phi_{\sigma,t}=(\lambda^2f^t(L(\lambda)))_+|_{\lambda=\lambda_\sigma}\Phi_\sigma.
\end{equation}
According to Theorem~\ref{thm:Ham}, these are Hamiltonian systems
with Hamiltonian functions
\begin{equation}\fl
   \begin{array}{l}
   \HH{22} H^x=\frac
   1{64}\tr\res\Bigg(\lambda\Big(L(\lambda)^4-6L(\lambda)^2\Big)\Bigg)\\
   \HH{22}\quad=-\frac 1{128}\tr\res\Bigg(\lambda\Big(L(\lambda)^6
   -5L(\lambda)^4+15L(\lambda)^2\Big)\Bigg)\\
   \HH{22}\quad=\innerp131-\innerp241+\big(\innerp120-\innerp340\big)^2
   +\frac 14\big(\innerp 110-\innerp 220+\innerp 330-\innerp 440\big)^2,\\
   \HH{22} H^t=\frac 1{64}\tr\res\Bigg(\lambda^2\theta
   \Big(L(\lambda)^6-5L(\lambda)^4+15L(\lambda)^2\Big)\Bigg)\\
   \HH{22}\quad=2(\innerp142+\innerp232)
   +2(\innerp121+\innerp341)(\innerp 110-\innerp 220+\innerp 330-\innerp
   440)\\
   \HH{22}\quad\quad-2(\innerp120-\innerp340)(\innerp 111-\innerp 221-\innerp
   331+\innerp441)\\
   \HH{22}\quad\quad+\!(\innerp140+\innerp230)
   \Big((\innerp 110-\innerp 220+\innerp 330-\innerp
   440)^2\!+\!4(\innerp120-\innerp340)^2\Big).
   \end{array}
\end{equation}

According to Theorem~\ref{thm:result} with $\theta_1=I$ and
$\theta_2=\theta$, these Hamiltonian systems are integrable in
Liouville sense. The solutions of the corresponding Hamiltonian
equations satisfy the nonlinear Schr\"odinger equation. This
recovers the results in~\cite{bib:ZRG}.

\subsection{$u(n)$ AKNS system}

Denote $I$ and $J$ as in the above subsection. The $x$ part of the
$u(n)$ AKNS system is the linear system
\begin{equation}
   \Phi_x=(\lambda K+P)\Phi.
   \label{eq:LPun}
\end{equation}
Here $K=(a_jJ\delta_{jk})_{1\le j,k\le n}$, $a_j$ $(j=1,\cdots,n)$
are real numbers such that $a_1,\cdots,a_n$ are distinct.
$P=(u_{jk}I+v_{jk}J)_{1\le j,k\le n}$ with $u_{jj}=v_{jj}=0$,
$u_{kj}=-u_{jk}$, $v_{kj}=v_{jk}$ $(j,k=1,\cdots,n)$.

Here we have written the $u(n)$ AKNS system in real form, which is
equivalent to usual complex form.

Now $m_1=2$, $N=1$, $\omega_1=1$, $\Omega_1=W=(-J\delta_{jk})_{1\le
j,k\le n}$. Then
\begin{equation}
   \begin{array}{l}
   \D\DDDD_0(=\DDDD_1)=\Big\{(a_{jk}I+b_{jk}J)_{1\le j,k\le
   n}\,|\,a_{jk},b_{jk}\in\hr,\\
   \D\qquad a_{kj}=-a_{jk},b_{kj}=b_{jk}\;
   (j,k=1,\cdots,n)\Big\},\\
   \D\D\DDDD_0\cap\ker\ad K=\Big\{(c_jJ\delta_{jk})_{1\le j,k\le n}
   \,|\,c_j\in\hr\;(j=1,\cdots,n)\Big\},\\
   \D\Theta_0=\{\pm I_{2n\times2n}\}.
   \end{array}
\end{equation}

Let $f^x=\zeta$ where $\zeta$ is given by Theorem~\ref{thm:red},
then $f^x(K)=K$.

Take $\kappa=-1$. Let $\Phi_\sigma$ $(\sigma=1,\cdots,r)$ be column
solutions of (\ref{eq:LPun}) with $\lambda=\lambda_\sigma$. By
(\ref{eq:Laxop}), the Lax matrix is
$L(\lambda)=(L_{jk}(\lambda))_{1\le j,k\le n}$ with
\begin{equation}\fl
   \begin{array}{l}
   \D L_{jk}(\lambda)=\left(\begin{array}{cc}0&-a_j\\a_j&0\end{array}\right)\delta_{jk}\\
   \D+\sum_{\sigma=1}^N\frac{1}{\lambda-\lambda_\sigma}\left(\begin{array}{cc}
   \phi_{2j-1,\sigma}\phi_{2k,\sigma}-\phi_{2j,\sigma}\phi_{2k-1,\sigma}
   &-\phi_{2j-1,\sigma}\phi_{2k-1,\sigma}-\phi_{2j,\sigma}\phi_{2k,\sigma}\\
   \phi_{2j-1,\sigma}\phi_{2k-1,\sigma}+\phi_{2j,\sigma}\phi_{2k,\sigma}
   &\phi_{2j-1,\sigma}\phi_{2k,\sigma}-\phi_{2j,\sigma}\phi_{2k-1,\sigma}\end{array}\right).
   \end{array}
\end{equation}
By Theorem~\ref{thm:Laxop} and \ref{thm:red}, the nonlinear
constraint is
\begin{equation}
   \begin{array}{l}
   \D u_{jk}=\innerp{2j-1}{2k}0-\innerp{2j}{2k-1}0,\quad
   \D v_{jk}=\innerp{2j-1}{2k-1}0+\innerp{2j}{2k}0\quad(j\ne k).
   \end{array}\label{eq:unconstr}
\end{equation}

Under this constraint, the Lax pair becomes a system of ODEs
\begin{equation}
   \Phi_{\sigma,x}=(\lambda
   \zeta(L(\lambda)))_+|_{\lambda=\lambda_\sigma}\Phi_\sigma.
   \label{eq:LPunb}
\end{equation}

It is too complicated to derive the Hamiltonian functions from
Theorem~\ref{thm:Ham} directly. However, it can be easily integrated
from (\ref{eq:LPunb}) to get the Hamiltonian functions since the
action of $\zeta$ is simply to remove the $\ker\ad K$ component from
$\widetilde P$. The result is that (\ref{eq:LPunb}) is a Hamiltonian
system with Hamiltonian function
\begin{equation}\fl
   \begin{array}{l}
   \D H^x=\frac 12\sum_{j=1}^n a_j(\innerp{2j-1}{2j-1}{1}+\innerp{2j}{2j}{1})\\
   \D\qquad+\frac{1}{4}\sum_{j,k=1\atop j\ne
   k}^n(\innerp{2j}{2k-1}{0}-\innerp{2j-1}{2k}{0})^2
   +\frac{1}{4}\sum_{j,k=1\atop j\ne
   k}^n(\innerp{2j-1}{2k-1}{0}+\innerp{2j}{2k}{0})^2.
   \end{array}
\end{equation}

\begin{remark}
This process is just the binary nonlinearization~\cite{bib:Ma} for
the $u(n)$ AKNS system~\cite{bib:MaZhou}. In fact, for any Lax pair
with unitary symmetry, the complex structure induces a natural
symplectic structure. Therefore, for any finite dimensional
Hamilto\-nian systems derived by nonlinearization method from the
$u(n)$ AKNS system, their conserved integrals, $r$ matrices and the
Liouville integrability are derived naturally from the results of
the present paper, although the Lax matrix and the Hamiltonian
functions may be derived more simply by direct computation.
\end{remark}

\begin{remark}
The nonlinear Schr\"odinger equation is also included in the $u(2)$
AKNS system. Hence the nonlinear constraint given here is also
applicable to the nonlinear Schr\"odinger
equation~\cite{bib:ZRGNLS}. However, it is different from that in
Subsection~6.2 because the symplectic structure here is derived
directly from the complex structure, while that in Subsection~6.2 is
the standard one in $sl(2,\hc)$ which is isomorphic to $sp(1,\hc)$.
\end{remark}

\subsection{$n$ wave equation}

The $n$-wave equation is the integrability condition of the Lax pair
\begin{equation}
   \Phi_x=(\lambda K+P)\Phi,\quad \Phi_t=(\lambda K'+Q)\Phi.
   \label{eq:LPnwave}
\end{equation}
Here $K=(a_jJ\delta_{jk})_{1\le j,k\le n}$,
$K'=(b_jJ\delta_{jk})_{1\le j,k\le n}$, $a_j,b_j$ $(j=1,\cdots,n)$
are real numbers such that $a_1,\cdots,a_n$ are distinct.
$P=(u_{jk}I+v_{jk}J)_{1\le j,k\le n}$ with $u_{jj}=v_{jj}=0$,
$u_{kj}=-u_{jk}$, $v_{kj}=v_{jk}$ $(j,k=1,\cdots,n)$. Moreover, $\D
Q=\Big(\frac{b_j-b_k}{a_j-a_k}(u_{jk}I+v_{jk}J)\Big)_{1\le j,k\le
n}$. Then $[K,Q]=[K',P]$. Clearly the $n$ wave equation is a special
equation in the $u(n)$ AKNS system. Hence we only need to consider
the $t$-part of the Lax pair.

For the $n$ wave equation, $N$, $m_1$, $\omega_1$, $\Omega_1=W$,
$\DDDD_0$, $\Theta$, the Lax matrix $L(\lambda)$ and the nonlinear
constraint (\ref{eq:unconstr}) are the same as those in the last
subsection for the $u(n)$ AKNS system.

Since $\det\Big((\sqrt{-1}a_j)^{k-1}\Big)_{1\le j,k\le n}\ne 0$, the
linear system
\begin{equation}
   \sum_{k=0}^{n-1}(a_jJ)^{n-k-1}(\alpha_kI+\beta_kJ)=b_jJ\quad
   (j=1,\cdots,n)
\end{equation}
has a unique real solution $\alpha_j$, $\beta_j$ $(j=1,\cdots,n)$. Let
\begin{equation}
   \hat f_j=((\alpha_jI+\beta_jJ)\delta_{ab})_{1\le a,b\le n},\quad
   \hat f(\tau)=\sum_{j=1}^n\hat f_{n-j}\tau^{j-1},
\end{equation}
then $\hat f(K)=K'$. Let $f^t=\hat f\circ\zeta$ where $\zeta$ is
given by Theorem~\ref{thm:red}, then $f^t(K)=K'$.

Under the constraint (\ref{eq:unconstr}), the Lax pair
(\ref{eq:LPnwave}) becomes two systems of ODEs
\begin{equation}
   \Phi_{\sigma,x}=(\lambda f^x(L(\lambda)))_+|_{\lambda=\lambda_\sigma}\Phi_\sigma,\quad
   \Phi_{\sigma,t}=(\lambda
   f^t(L(\lambda)))_+|_{\lambda=\lambda_\sigma}\Phi_\sigma.
   \label{eq:LPnwaveb}
\end{equation}

Expand $\widetilde L(\lambda)=\zeta(L(\lambda))$ as
\begin{equation}
   \begin{array}{l}
   \D \widetilde L=K+\lambda^{-1} \widetilde P+o(\lambda^{-1}),\\
   \D \widetilde L^k=K^k+\lambda^{-1}\sum_{j=0}^{k-1}K^j\widetilde PK^{k-j-1}+o(\lambda^{-1}).
   \end{array}
\end{equation}
With the identity
\begin{equation}
   \frac{b_\mu-b_\nu}{a_\mu-a_\nu}I=\sum_{k=1}^n\sum_{j=0}^{k-2}
   (\alpha_{n-k}I+\beta_{n-k}J)(a_\mu J)^j(a_\nu J)^{k-j-2},
\end{equation}
we have
\begin{equation}\fl
   \begin{array}{l}
   \D f^t(L)_{\mu\nu}=\hat f(\widetilde L)_{\mu\nu}\\
   \D=K'_{\mu\nu}+\lambda^{-1}\sum_{k=1}^n\sum_{j=0}^{k-2}
   (\alpha_{n-k}I+\beta_{n-k}J)(a_\mu J)^j\widetilde P_{\mu\nu}(a_\nu J)^{k-j-2}
   +o(\lambda^{-1})\\
   \D=K'_{\mu\nu}+\frac{b_\mu-b_\nu}{a_\mu-a_\nu}\widetilde P_{\mu\nu}.
   \end{array}
\end{equation}
This gives the constraint on $Q$:
$\D\widetilde Q_{\mu\nu}=\frac{b_\mu-b_\nu}{a_\mu-a_\nu}\widetilde
P_{\mu\nu}$.

By integration, (\ref{eq:LPnwaveb}) becomes Hamiltonian systems with
Hamiltonian functions
\begin{equation}\fl
   \begin{array}{l}
   \D H^x=\frac 12\sum_{j=1}^n a_j(\innerp{2j-1}{2j-1}{1}+\innerp{2j}{2j}{1})\\
   \D\qquad+\frac{1}{4}\sum_{j,k=1\atop j\ne
   k}^n(\innerp{2j}{2k-1}{0}-\innerp{2j-1}{2k}{0})^2
   +\frac{1}{4}\sum_{j,k=1\atop j\ne
   k}^n(\innerp{2j-1}{2k-1}{0}+\innerp{2j}{2k}{0})^2,\\
   \D H^t=\frac 12\sum_{j=1}^n b_j(\innerp{2j-1}{2j-1}{1}+\innerp{2j}{2j}{1})\\
   \D\qquad+\frac{1}{4}\sum_{j,k=1\atop j\ne
   k}^n\frac{b_j-b_k}{a_j-a_k}(\innerp{2j}{2k-1}{0}-\innerp{2j-1}{2k}{0})^2
   +\frac{1}{4}\sum_{j,k=1\atop j\ne
   k}^n\frac{b_j-b_k}{a_j-a_k}(\innerp{2j-1}{2k-1}{0}+\innerp{2j}{2k}{0})^2.
   \end{array}
\end{equation}

These are involutive Hamiltonian systems which are integrable in
Liouville sense. Each solution of these Hamiltonian systems gives a
solution of the $n$ wave equation~\cite{bib:MaZhou}.

\section{Two dimensional hyperbolic $C_n^{(1)}$ Toda
equation}\label{sect:Cn}

The two dimensional hyperbolic $C_n^{(1)}$ Toda equation is
\begin{equation}
   \begin{array}{l}
   u_{1,xt}=\E{2u_1}-\E{u_2-u_1},\quad
   u_{n,xt}=\E{u_{n}-u_{n-1}}-\E{-2u_{n}},\\
   u_{j,xt}=\E{u_j-u_{j-1}}-\E{u_{j+1}-u_j}\quad
   (2\le j\le n-1).
   \end{array}\label{eq:evolevenoddn}
\end{equation}
It has a Lax pair
\begin{equation}
   \D\Phi_x=(\lambda K+P(x,t))\Phi,\quad
   \D\Phi_t=\lambda^{-1}Q(x,t)\Phi
   \label{eq:LPC}
\end{equation}
where $K=(\delta_{j+1,k})_{1\le j,k\le 2n}$,
$P=(p_j\delta_{jk})_{1\le j,k\le 2n}$ with $p_j=u_{j,x}$ for
$j=1,\cdots,n$ and $p_j=-u_{2n+1-j,x}$ for $j=n+1,\cdots,2n$,
$Q=(q_k\delta_{j,k+1})_{1\le j,k\le 2n}$ with $q_k=\E{u_{k+1}-u_k}$
for $k=1,\cdots,n-1$, $q_n=\E{-2u_n}$, $q_k=\E{u_{2n+1-k}-u_{2n-k}}$
for $k=n+1,\cdots,2n-1$, $q_{2n}=\E{2u_1}$. Note that
$p_j+p_{2n+1-j}=0$, $q_j=q_{2n-j}$ and $q_1q_2\cdots q_{2n}=1$. Here
we use the convention $q_{2n+j}=q_j$ etc.

Written in components, (\ref{eq:LPC}) is
\begin{equation}
   \phi_{j,x}=\lambda\phi_{j+1}+p_j\phi_j,\quad \phi_{j,t}=\lambda^{-1} q_{j-1}\phi_{j-1}.
\end{equation}
(\ref{eq:evolevenoddn}) is equivalent to
\begin{equation}
   Q_x=[P,Q],\quad P_t+[K,Q]=0,
\end{equation} or
\begin{equation}
   q_{k,x}=(p_{k+1}-p_k)q_k,\quad p_{k,t}=q_{k-1}-q_k,
\end{equation}
which are
equivalent to the integrability condition of (\ref{eq:LPC}).

Now $N=1$, $m_1=2n$, $\D\omega_1=\omega=\rho^2$ where
$\D\rho=\exp\Big(\frac{\pi\I}{2n}\Big)$,
$W=((-1)^j\delta_{j,2n+1-k})_{1\le j,k\le 2n}$,
$\Omega_1=(\rho^{-2j+1}\delta_{jk})_{1\le j,k\le 2n}$. Then
\begin{equation}
   \begin{array}{l}
   \D\DDDD_k=\{(a_{ij})_{2n\times 2n}\,|\,a_{ij}\ne 0\hbox{ only when }j-i\equiv k\mod 2n,\\
   \qquad\hbox{ and satisfy }(-1)^ka_{i,i+k}+a_{1-k-i,1-i}=0\,(1\le i\le
   2n)\},\\
   \DDDD_0\cap\ker\ad K=\{0\},\\
   \Theta_0=\{\pm I_{2n\times 2n}\,|\,c\in\hr\},\quad
   \Theta_k=\{0\}\quad(k=1,2,\cdots,2n-1).
   \end{array}
\end{equation}
We have $P\in\DDDD_0$, $Q\in\DDDD_{-1}$.

Take $\D\kappa=\frac 1{2n}$, Let $\Phi_\sigma$ $(\sigma=1,\cdots,r)$
be column solutions of (\ref{eq:LPC}) with $\lambda=\lambda_\sigma$.
By (\ref{eq:Laxop}), the Lax matrix is
\begin{equation}
   L(\lambda)=
   K+\frac 1{2n}\sum_{\alpha=1}^{2n}\sum_{\sigma=1}^r
   \frac{\Omega^\alpha \Phi_\sigma
   \Phi_\sigma^T(\Omega^{\alpha})^TW}
   {\lambda-\omega^\alpha\lambda_\sigma},
\end{equation}
whose entries are
\begin{equation}
   L_{jk}(\lambda)=K_{jk}+\sum_{\sigma=1}^{r}
   \frac{(-1)^{k-1}\lambda_\sigma^{\{j-k\}}\lambda^{2n-1-\{j-k\}}}
   {\lambda^{2n}-\lambda_\sigma^{2n}}
   \phi_{j\sigma}\phi_{2n+1-k,\sigma},\label{eq:Cn_Ljk}
\end{equation}
where $\{k\}$ is the remainder of $k$ divided by $2n$. Here we have
used the identity
\begin{equation}
   \sum_{\alpha=0}^{2n-1}\frac{\omega^{-p\alpha}}{\lambda-\omega^\alpha\lambda_\sigma}
   =\frac{2n\lambda_\sigma^{\{p\}}\lambda^{2n-1-\{p\}}}{\lambda^{2n}-\lambda_\sigma^{2n}}.
\end{equation}
(see Lemma~2 in~\cite{bib:ZhouHam}).

\begin{theorem}
Under the constraint
\begin{equation}
   \E{u_j}=\Gamma^{j-\frac 12}\gamma_0^{-\frac 12}\prod_{k=1}^{j-1}\gamma_k^{-1}
   \label{eq:TodaCu}
\end{equation}
where
\begin{equation}
   \gamma_j=1-(-1)^j\innerp{j}{-j}{-1},\quad
   \Gamma=\prod_{k=1}^{2n}\gamma_k^{\frac 1{2n}}, \label{eq:Gamma}
\end{equation}
the Lax pair (\ref{eq:LPC}) of the two dimensional hyperbolic
$C_n^{(1)}$ Toda equation is changed to a system of ODEs
\begin{equation}\fl
   \phi_{j\sigma,x}=\lambda_\sigma\phi_{j+1,\sigma}
   +(-1)^{j-1}\innerp{j}{1-j}{0}\phi_{j,\sigma},\quad
   \phi_{j\sigma,t}=\lambda_\sigma^{-1}\prod_{k=1}^{2n}\gamma_k^{\frac 1{2n}}
   \gamma_{j-1}^{-1}\phi_{j-1,\sigma},
   \label{eq:TodaCconstrLP0}
\end{equation}
or equivalently,
\begin{equation}\fl
   \Phi_{\sigma,x}=(\lambda L(\lambda))_+|_{\lambda=\lambda_\sigma}\Phi_\sigma,\quad
   \Phi_{\sigma,t}=\lambda_\sigma^{-1}\Big(\frac 1{2n}\tr(L(0)^{2n})\Big)
   ^{\frac 1{2n}-1}L(0)^{2n-1}\Phi_\sigma.
   \label{eq:TodaCconstrLP}
\end{equation}
These ODEs are Liouville integrable Hamiltonian systems with the
Hamiltonian func\-tions
\begin{equation}
   \begin{array}{l}
   \D H^x=\frac 12\tr\res\Big(\lambda L^2(\lambda)\Big)=\sum_{j=1}^{2n}(-1)^j\innerp j{2n+2-j}1
   +\frac 12\sum_{j=1}^{2n}(\innerp j{2n+1-j}0)^2,\\
   \D H^t=-n\Big(\frac 1{2n}\tr(L(0)^{2n})\Big)^{\frac 1{2n}}
   =-n\prod_{j=1}^{2n}\Big(1-(-1)^{j-1}\innerp{j}{-j}{-1}\Big)^{\frac{1}{2n}}.
   \end{array}\label{eq:TodaCHams}
\end{equation}
\end{theorem}

\begin{demo}
Take $f^x(\tau)=\tau$. According to Theorem~\ref{thm:Laxop}, the
nonlinear constraint is
\begin{equation}
   p_j=(-1)^{j-1}\innerp{j}{1-j}0. \label{eq:nonlconsC}
\end{equation}
We should mention that (\ref{eq:TodaCu}) is compatible with
(\ref{eq:nonlconsC}) under the relation $p_j=u_{j,x}$. In fact, by
the definition of $\gamma_j$ and the constraint
(\ref{eq:nonlconsC}),
\begin{equation}
   -\frac{\gamma_{j,x}}{\gamma_j}=p_{j+1}-p_j.
\end{equation}
Hence $\Gamma_x=0$. (This can also be obtained from
(\ref{eq:TodaCtrL0}) below.) Then from (\ref{eq:TodaCu}),
\begin{equation}
   u_{j,x}=-\frac 12\frac{\gamma_{0,x}}{\gamma_0}
   -\sum_{k=1}^{j-1}\frac{\gamma_{k,x}}{\gamma_k}=p_j
\end{equation}
with the relation $p_1+p_0=0$.

Under the constraint (\ref{eq:nonlconsC}), the first equation of the
Lax pair (\ref{eq:LPC}) becomes
\begin{equation}
   \Phi_{\sigma,x}=(\lambda L(\lambda))_+|_{\lambda=\lambda_\sigma}\Phi_\sigma
   =(\lambda_\sigma K+\widetilde P)\Phi_\sigma
\end{equation}
where $\widetilde P=((-1)^{j-1}\innerp{j}{1-j}0)_{1\le j,k\le 2n}$.

According to Theorem~\ref{thm:Ham}, this is a Hamiltonian system
with Hamiltonian function
\begin{equation}\fl
   H^x=\frac 12\tr\res\Big(\lambda L^2(\lambda)\Big)
   =\sum_{j=1}^{2n}(-1)^j\innerp j{2n+2-j}1
   +\frac 12\sum_{j=1}^{2n}(\innerp j{2n+1-j}0)^2.
\end{equation}

The coefficient of the second equation of the Lax pair (\ref{eq:LPC}) is not a
polynomial of $\lambda$. Hence we can not use the above general method and
should construct its nonlinear constraint and Hamiltonian function
directly. Similar to (\ref{eq:monox}), we have
\begin{equation}
   (\Omega^\alpha \Phi_\sigma \Phi_\sigma^T(\Omega^\alpha)^TW)_t
   =\frac 1{\omega^\alpha\lambda_{\sigma}}
   [Q,\Omega^\alpha \Phi_\sigma \Phi_\sigma^T(\Omega^\alpha)^TW)].
\end{equation}
Hence
\begin{equation}
   L_t-\frac 1{\lambda}[Q,L]
   =-\frac1{\lambda}[Q,K-\hat K]
\end{equation}
where
\begin{equation}
   \begin{array}{l}
   \D\hat K=(\hat K_{jk})=\frac{1}{2n}\sum_{\sigma=1}^r
   \sum_{\alpha=0}^{2n-1}(\omega^\alpha\lambda_\sigma)^{-1}
   \Omega^\alpha \Phi_\sigma \Phi_\sigma^T(\Omega^\alpha)^TW,\\
   \hat K_{jk}
   \D=(-1)^j\sum_{\sigma=1}^r\lambda_\sigma^{-1}
   (\Phi_\sigma \Phi_\sigma^T)_{j,-j}\delta_{j,k-1}=(1-\gamma_j)\delta_{j,k-1}.
   \end{array}
\end{equation}
$[Q,K-\hat K]=0$ holds if and only if
$\gamma_jq_j=\gamma_{j+1}q_{j+1}$. This is equivalent to
\begin{equation}
   q_j=\gamma_j^{-1}\widetilde \Gamma \label{eq:qbyGamma}
\end{equation}
for certain function $\widetilde \Gamma$. However, since
$q_1q_2\cdots q_{2n}=1$, we have
\begin{equation}
    \widetilde \Gamma=\Big(\prod_{k=1}^{2n}\gamma_k\Big)^{\frac 1{2n}}=\Gamma,
\end{equation}
and the nonlinear constraint becomes
\begin{equation}
   q_j=\Gamma\gamma_j^{-1},
\end{equation}
which is equivalent to (\ref{eq:TodaCu}). Meanwhile, the second equation
of the Lax pair (\ref{eq:LPC}) can be written as the second equation of
(\ref{eq:TodaCconstrLP0}).

From (\ref{eq:Cn_Ljk}), we have
\begin{equation}
   \begin{array}{rl}
   (L(0))_{jk}
   =&\D \delta_{j+1,k}-\sum_{\sigma=1}^r
   (-1)^j\lambda_\sigma^{-1}\phi_{j\sigma}\phi_{-j,\sigma}
   \delta_{j+1,k}=\gamma_j\delta_{j+1,k}.
   \end{array}
\end{equation}
Hence
\begin{equation}
   \tr(L(0))^{2n}=2n\prod_{k=1}^{2n}\gamma_k
   \label{eq:TodaCtrL0}
\end{equation}
and
\begin{equation}
   \Big(L(0)^{2n-1}\Big)_{jk}=\prod_{l=1\atop l\ne
   j-1}^{2n}\gamma_l\delta_{j,k+1}.
\end{equation}

With the constraint (\ref{eq:TodaCu}), the second equation of
(\ref{eq:TodaCconstrLP0}) can be written as the second equation of
(\ref{eq:TodaCconstrLP}).

With $H^t$ in (\ref{eq:TodaCHams}), we have
\begin{equation}\fl
   \begin{array}{l}
   \D\sum_{k=1}^{2n}\hat W_{jk}\frac{\partial H^t}{\partial\phi_{k\sigma}}
   =\frac 12\sum_{k,a,b=1}^{2n}(-1)^{j}\delta_{j+k,1}
   \Big(\frac 1{2n}\tr L(0)^{2n}\Big)^{\frac 1{2n}-1}
   \Big(L(0)^{2n-1}\Big)_{ba}\frac{\partial L(0)_{ab}}{\partial\phi_{k\sigma}}\\
   \D=\frac 12\sum_{k,a,b=1}^{2n}\sum_{\sigma=1}^r
   (-1)^{j-1}\delta_{j+k,1}\Big(\frac 1{2n}\tr L(0)^{2n}\Big)^{\frac 1{2n}-1}
   \Big(L(0)^{2n-1}\Big)_{ba}\lambda_\sigma^{-1}\\
   \D\qquad\cdot(-1)^a(\delta_{ak}\phi_{-a,\sigma}\delta_{a+1,b}
   +\phi_{a\sigma}\delta_{-a,k}\delta_{a+1,b})\\
   \D=\frac 12\sum_{k=1}^{2n}\sum_{\sigma=1}^r(-1)^{j+k-1}\lambda_\sigma^{-1}
   \Big(\frac 1{2n}\tr L(0)^{2n}\Big)^{\frac
   1{2n}-1}\\
   \D\quad\cdot\Big((L(0)^{2n-1})_{k+1,k}+(L(0)^{2n-1})_{1-k,-k}\Big)
   \phi_{-k,\sigma}\delta_{j+k,1}\\
   \D=\lambda_\sigma^{-1}\frac{\D\Big(\frac 1{2n}\tr L(0)^{2n}\Big)^{\frac 1{2n}}}
   {\D 1-(-1)^{j-1}\innerp{j-1}{1-j}{-1}}\phi_{j-1,\sigma}.
   \end{array}
\end{equation}
Hence $H^t$ is the Hamiltonian function of the second equation of
(\ref{eq:TodaCconstrLP}).

According to Theorem~\ref{thm:invol} and \ref{thm:indep}, the
Hamiltonian systems given by both $H^x$ and $H^t$ are Liouville
integrable. The theorem is proved.
\end{demo}

Therefore, any solution of the integrable Hamiltonian systems with
Hamiltonian functions (\ref{eq:TodaCHams}) gives a solution of the
two dimensional hyperbolic $C_n^{(1)}$ Toda equation. The
corresponding symplectic structure is the natural one of
$C_n^{(1)}$. The Hamiltonian systems (\ref{eq:TodaCHams}) are
simpler than (with space of lower dimension) that presented
in~\cite{bib:ZhouHam} where the symplectic structure is derived from
the complex structure.

\section*{Acknowledgements} This work was supported by the
National Basic Research Program of China (973 Program)
(2007CB814800) and the Key Laboratory of Mathematics for Nonlinear
Sciences of Ministry of Education of China. The author is grateful
to Prof.\ Ruguang Zhou and Prof.\ Shenglin Zhu for helpful
discussions.

\end{document}